\documentclass{article}

\usepackage[a4paper,margin=2.5cm]{geometry}
\usepackage{graphicx}
\usepackage{multicol,multirow}

\usepackage{amsmath}
\usepackage{amsthm}
\usepackage{bm}

\usepackage{mathrsfs} 

\usepackage{apacite}
\usepackage{rotating}
\usepackage{appendix}
\usepackage[authoryear]{natbib}

\usepackage[T1]{fontenc}
\usepackage{times}
\usepackage{sourcesanspro}
\usepackage{newtxmath}

\usepackage{textcomp}
\usepackage{xcolor}
\usepackage{hyperref}
\usepackage{booktabs}
\usepackage{array}

\usepackage{algorithm}
\usepackage{algorithmic}
\usepackage{subcaption}

\usepackage{pdfcomment}
\usepackage{changes}
\usepackage{todonotes}

\newcommand{\keywords}[1]{\par\noindent\textbf{Keywords: }#1}

\title{Bayesian Tendon Breakage Localization under Model Uncertainty Using Distributed Fiber Optic Sensors}

\author{Daniel Andrés Arcones$^{a,b,*}$, Aeneas Paul$^{c,*}$,\\ Martin Weiser$^{d}$, David Sanio$^{c}$, Peter Mark$^{c}$, Jörg F. Unger$^{b}$\\
	\small $^{a}$Technical University of Munich, Garching bei München, Germany \\
	\small $^{b}$Bundesanstalt für Materialforschung und -prüfung, Berlin, Germany \\
	\small $^{c}$Ruhr University Bochum, Bochum, Germany \\
	\small $^{d}$Zuse Institute Berlin, Berlin, Germany\\\\
	\small
	\begin{tabular}{@{}l l@{}}
		$^{*}$Corresponding authors: &
		Daniel Andrés Arcones; \tt{daniel.andres-arcones@bam.de} \\
		& Aeneas Paul; \tt{aeneas.paul@ruhr-uni-bochum.de}
	\end{tabular}
}
\date{08 April 2026}

\begin{document}
	\maketitle
\begin{abstract}
	This study develops a Bayesian, uncertainty-aware framework for tendon breakage localization in pre-stressed concrete members using high-resolution data from distributed fiber-optic sensors (DFOS). DFOS enable full-field monitoring of strain changes on the surface of pre-stressed concrete members due to such failure. A finite element model (FEM) of an experimental tendon-breakage test is constructed, and model parameters are calibrated probabilistically against DFOS measurements. To capture model-form uncertainty (MFU), stochastic perturbations are embedded directly into material parameters, enabling the joint inference of physical properties and MFU within a unified probabilistic framework. Gaussian Process surrogates are employed to efficiently emulate the nonlinear FEM response, supporting computationally tractable Bayesian inference. A $\phi$-divergence-based influence analysis identifies the DFOS measurements that most strongly shape the posterior distributions, providing interpretable diagnostics of sensor informativeness and model adequacy. The calibrated parameters and embedded uncertainties are then transferred to a FEM of a full-scale structural configuration, enabling prediction of tendon breakage localization under realistic conditions. A separability analysis of the predictive strain distributions quantifies the identifiability of tendon breakage at varying depths, assessing the confidence with which different damage scenarios can be distinguished given the propagated uncertainties. Results demonstrate that the framework achieves robust parameter calibration, interpretable diagnostics, and uncertainty-informed damage detection, integrating experimental data, embedded MFU, and probabilistic modeling. By systematically propagating both experimental and model uncertainties, the approach supports reliable tendon breakage localization, informed decision-making, and optimal DFOS placement.
\end{abstract}

\section{Introduction}
\label{sec1}

Understanding the location of tendon breakage in pre-stressed concrete members is essential for assessing structural integrity \citep{Sieradzki.1987}, guiding emergency interventions, and designing effective monitoring strategies within Structural Health Monitoring (SHM) \citep{Bergmeister.2015b, Farrar.2007, Richter.2025}. In last years, SHM approaches have therefore gained popularity for extending the life of critical structures such as bridges \citep{Becks2024, Herbers2024, Kang2025}. In such cases, however, destructive experiments cannot be performed, making direct calibration or validation of mechanical models against in-situ tendon failures infeasible \citep{Pirskawetz.2023}. As a result, model calibration must rely on controlled laboratory experiments, where tendon breakage can be safely induced and instrumented. The central challenge is therefore not only to calibrate a model to experimental data, but to do so in a way that allows reliable propagation of uncertainties from laboratory-scale tests to simulations of real structural systems, where tendon breakage localization is ultimately required.

In particular, tendon breakage produces abrupt and highly localized redistribution of strains that propagate along the member. These strain changes can be captured using a spatial grid of distributed fiber-optic sensors (DFOS) \citep{Paul2024}, which provide high-resolution, full-field strain change profiles \citep{Barrias.2016}. Such data contain rich information about load transfer mechanisms and bond–slip behavior, making DFOS a powerful sensing technology for tendon-related damage assessment. However, leveraging dense DFOS measurements for tendon breakage localization requires uncertainty-aware methodologies capable of reconciling measurement noise, nonlinear physics, and modeling errors, particularly when results must be transferred beyond the experimental configuration.

A natural framework for simulating the tendon–concrete interaction \citep{Ayoub2010,Abdelatif.2017} and the resulting strain redistribution after breakage \citep{Seiffert2019,Paul.2025} are provided by Finite element models (FEM). Yet, even detailed nonlinear FEMs are affected by model-form uncertainty (MFU) due to idealized constitutive laws, uncertain boundary conditions, and geometric simplifications. When calibration is performed solely against laboratory data, these deficiencies can lead to biased or overconfident predictions if not explicitly accounted for. Discrepancy-based approaches such as the Kennedy–O’Hagan (KOH) formulation \citep{Kennedy2001} can improve calibration fidelity in the experimental configuration \citep{AndresArcones2024}, but the inferred discrepancy terms are tied to the calibration setup and do not naturally transfer to new structural configurations, limiting their usefulness for real-world tendon breakage localization.

To overcome this limitation, embedded model-form uncertainty is adopted, whereby uncertainty is absorbed directly into an enlarged stochastic parameterization of the model [\citealt{Sargsyan2015}, \citealt{ Sargsyan2019}]. For tendon-breakage problems, sensitivity studies indicate that the Young’s modulus of the concrete matrix plays a dominant role in controlling the strain response \citep{Paul.2026}. Treating this modulus as a stochastic variable allows MFU to be represented within the constitutive behavior itself, enabling Bayesian inference approaches  to jointly estimate physical parameters and model uncertainty. This embedded representation produces probabilistic parameter descriptions that can be meaningfully transferred from laboratory experiments to simulations of real structural systems, without relying on non-transferable discrepancy corrections.

Bayesian calibration involving both deterministic and stochastic parameters requires repeated model evaluations, which are computationally prohibitive for detailed nonlinear FEMs. Gaussian Process (GP) surrogates are therefore employed to emulate FEM responses with quantified predictive uncertainty \citep{Rasmussen2006}, enabling efficient assimilation of full-field DFOS data and uncertainty propagation. Beyond calibration, the framework supports diagnostic analyses. A $\phi$-divergence-based influence analysis \citep{Weiss1996a} identifies which DFOS measurements most strongly affect posterior distributions, providing insight into sensor informativeness and model adequacy. This data-centric diagnostic capability is essential for improving physical models in complex structural systems.

Finally, the embedded stochastic parameters can be transferred into new FEM models representing more realistic structural configurations, such as T beam cross-sections or varying tendon embedment depths. This enables the prediction of strain fields in scenarios where no experimental data are available. In particular, comparing predictive distributions across different tendon positions allows the assessment of \emph{separability}, a measure of how distinguishable different structural configurations, realized by varying embedment depths of the tendons, are under the propagated uncertainty. This is also commonly known as structural identifiability \citep{Bellman1970}. Regions where predictive distributions overlap indicate sensor locations that provide limited information under the current propagated uncertainty, whereas regions with clear separation identify locations where measurements would be most informative for detecting tendon-related anomalies. This can be used for guiding potential sensor placement \citep{Ostachowicz2019}.

In this work, we present a fully uncertainty-aware workflow that integrates DFOS measurements, FEM modeling, Bayesian inference with embedded MFU, influence diagnostics, and uncertainty transfer from laboratory experiments to full scale structural simulations. The framework addresses the fundamental SHM challenge of localizing tendon breakage in systems where direct experimental calibration is impossible, providing a rigorous and transferable basis for uncertainty-informed structural assessment.

\section{Methodology}
\label{sec2}

\begin{figure}[h]
\centering
\includegraphics[width=0.9\textwidth]{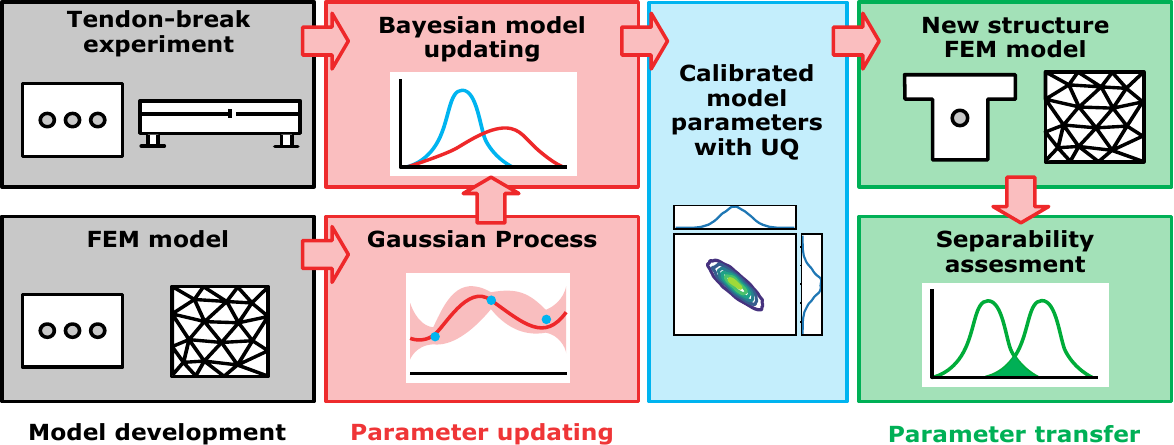}
\caption{Workflow for the parameter updating and transfer of the simulation of tendon breakage in pretensioned concrete.}
\label{fig:workflow}
\end{figure}

The overall workflow used in this study is summarized in Figure~\ref{fig:workflow}. The methodology combines surrogate modeling, Bayesian inference, uncertainty quantification, and influence analysis in a unified framework. First, a computational model of the experimental setup is developed. It is approximated using a Gaussian Process (GP) regressor in order to alleviate the cost of repeated high-fidelity simulations (Section~\ref{sec2.3}). Using observational data together with the GP surrogate, the model parameters are inferred through a Bayesian calibration procedure that explicitly accounts for model-form uncertainty (Section~\ref{sec2.4}). Based on the resulting uncertainty estimates, an influence analysis is performed to assess the contribution of different data subsets and parameters to the inferred posterior distributions (Section~\ref{sec2.5}). This analysis provides insight into dominant sources of discrepancy and supports targeted model improvements. Finally, the inferred parameters and their associated uncertainties are propagated to related prediction tasks in order to assess the robustness of the results under uncertainty (Section~\ref{sec2.6}). The application of this methodology to the specific case study of tendon breakage in reinforced concrete is presented in a dedicated section, where the experimental setup, the high-fidelity model, and the corresponding results are described in detail.

\subsection{Bayesian Parameter Updating under Model Form Uncertainty}
\label{sec2.4}
\subsubsection{Model calibration}
We define the simulation-based forward model as
\begin{equation}
f : \mathcal{X}_f\times\Theta \subset \mathbb{R}^d  \to \mathbb{R}^m, \quad (\mathbf{x},\bm{\theta}) \mapsto f(\mathbf{x},\bm{\theta}),
\label{eq:def_forward_model}
\end{equation}
where $(\mathbf{x},\bm{\theta})\in\mathcal{X}_f\times\Theta \subset \mathbb{R}^d $ is a $d$-dimensional vector of input parameters composed by observation coordinates $\mathbf{x}$ and model parameters $\bm{\theta}$ and $f(\bm{\xi})\in\mathbb{R}^m$ is the corresponding $m$-dimensional simulation output. For calibration, we have the observations dataset $\mathcal{Y}=(\mathbf{x}_i,y_i|i=1,...,N)$ or in matrix form $\mathcal{Y}=(X,\mathbf{y})$. Each observation relates to the simulation model $f$ as
\begin{equation}
    y=f(x,\bm{\theta})+\delta_f(x)+\varepsilon_N,
\end{equation}
where $\delta_f(\mathbf{x})$ is a discrepancy term between observations and predictions of $f$ due to the model form uncertainty and $\varepsilon_N$ is some prescribed noise, which will be treated as $\varepsilon\sim\mathcal{N}(0,\sigma_\varepsilon^2)$.

Without MFU, the classical parameter updating problem consists in solving
\begin{equation}
    \bm{\theta}^*=\arg\min_{\bm{\theta}} \,\|\mathbf{y}-f(X,\bm{\theta})\|,
\end{equation}
where $\|\cdot\|$ is a norm to be defined, typically the Euclidean norm that leads to solving a minimization of the mean-squared error (MSE). Following the common principles of modularization \citep{Bayarri2007}, the training of the surrogate and the calibration of the parameters is performed sequentially and independently.

In the Bayesian framework, the main objective is to find the \textit{posterior} probability distribution of $\bm{\theta}^*$ given the observations. We start by disregarding the discrepancy term $\left(\delta_{f}=0\right)$ and imposing an error structure for the observations vector $\mathbf{y}$ such that it follows a multivariate normal distribution as $\mathbf{y}-f(X,\bm{\theta})\sim\mathcal{MVN}(\mathbf{0},\Sigma_\varepsilon)$, where $\Sigma_\varepsilon$ is the covariance matrix equal to $\sigma_\varepsilon^2I$ for independent normal errors. We obtain then the \textit{likelihood} distribution as
\begin{equation}
    \mathcal{L}(\bm{\theta):=\pi(\mathbf{y}|\bm{\theta})}=(2\pi)^{-\frac{N}{2}}\det\left(\Sigma_{\varepsilon}\right)^{-\frac{1}{2}}\exp\left(-\frac{1}{2}\|\mathbf{y}-f(X,\bm{\theta})\|^2_{\Sigma^{-1}_\varepsilon}\right),
    \label{eq:likelihood_gaussian}
\end{equation}
where $\|\mathbf{y}-f(X,\bm{\theta})\|^2_{\Sigma^{-1}_\varepsilon}$ denotes the euclidean norm of the residuals weighted by the inverse of the covariance of the noise model. One option is directly substituting the model $f(X,\bm{\theta})$ by the mean of its approximation $\mathbf{m}_{\mathrm{GP}}(\bm{\theta})=\mathbb{E}\left[\tilde{f}(X,\bm{\theta})\right]$, where $\tilde{f}(X,\bm{\theta})$ is the GP regression fitted for $f(X,\bm{\theta})$. The details on the GP implementation are delayed to Section \ref{sec2.3}.

Applying Bayes' theorem, it is possible to obtain the posterior probability distribution of $\bm{\theta}$ from the likelihood $\mathcal{L}(\bm{\theta})$ and the prior $\pi(\bm{\theta})$. In particular,
\begin{equation}    \pi(\bm{\theta}|\mathbf{y})\propto\pi(\mathbf{y}|\bm{\theta})\pi(\bm{\theta})=\mathcal{L}(\bm{\theta})\pi(\bm{\theta}).
\label{eq:bayes_theorem}
\end{equation}
Sampling approaches such as those based on Monte Carlo-Markov Chains (MCMC) draw realizations of the posterior distribution by evaluating samples of $\bm{\theta}$ at Equation \ref{eq:bayes_theorem}.

\subsubsection{Model form uncertainty}
Traditional Bayesian approaches do not treat model-form uncertainty (MFU) effectively, as the discrepancy term $\delta_f(x)$ is generally disregarded. Under additive Gaussian observation noise with diagonal covariance, Bayesian calibration reduces to a (weighted) least squares problem, which in the present case of a scalar diagonal covariance coincides with ordinary least squares. As a result, the inferred posterior distribution $\pi(\bm{\theta}\mid\mathbf{y})$ concentrates around the least squares estimate as the noise level decreases, and any unexplained variability in the observations is necessarily attributed to parameter uncertainty. Consequently, the posterior predictive distribution $f(X,\bm{\theta}\mid\mathbf{y})$ fails to reflect variability arising from structural model inadequacy, leading to overconfident predictions that do not capture the observed variability \citep{AndresArcones2024}.

One popular alternative is the framework from \cite{Kennedy2001}, that implements $\delta_f(x)$ as a flexible parametrized function that compensates the discrepancy, generally a GP. However, the $\delta_f(x)$ inferred within that framework cannot be propagated to other systems or datasets than the one used for calibration. A promising alternative is the embedding of the uncertainty within the parameter formulation, as proposed in \cite{Sargsyan2019}. The model with discrepancies is then reformulated as
\begin{equation}
    y = f(x,\bm{\theta}) + \delta_f(x) + \varepsilon_N
    \approx f\left(x,\bm{\tilde{\theta}}\right) + \varepsilon_N,
\end{equation}
where $\bm{\tilde{\theta}}$ is a stochastic extension of the parameter vector $\bm{\theta}$, endowed with a probability density $\pi_{\bar{\theta}}\left(\bm{\tilde{\theta}}\right)$ defined on the parameter space $\Theta$. This formulation implicitly accounts for the model discrepancy $\delta_f(x)$ through variability in the extended parameters. Introducing a probability distribution for $\bm{\tilde{\theta}}$ follows a hierarchical Bayes construction and prevents excessive concentration of the posterior distribution of the affected parameters and their associated predictions. The resulting predictive uncertainty can be calibrated to provide a suitable representation of the model-form uncertainty, provided that the extended model is sufficiently flexible to cover the range of the observations through variations in $\bm{\tilde{\theta}}$. The hyperparameters governing the density $\pi_{\bar{\theta}}$ are inferred jointly with the original model parameters. In this paper, we will follow the methodology developed in \cite{AndresArcones2024a}, defining explicitly such probability distributions as normal or log-normal, and adding their parameters to the inference. Therefore, $\bm\theta_\mathrm{ext}=\lbrace\bm{\theta}, \bm{\theta_\delta}\rbrace$, where $\bm{\theta_\delta}$ represents the vector of parameters associated with the stochastic extension for the discrepancy $\tilde{\bm{\theta}}$.

\subsubsection{Uncertainty propagation}
As $\tilde{\bm{\theta}}$ is a random variable, $f\left(x,\tilde{\bm{\theta}}\right)$ is also stochastic, which means that the probability distribution $\pi_{\tilde{\bm{\theta}}}$ of $\tilde{\bm{\theta}}$ must be either sampled or propagated through $f$ for its full description. As it is common in embedded approaches -- in contrast to hierarchical Bayes -- we opt for propagating $\tilde{\bm{\theta}}$ by using a Polynomial Chaos Expansion (PCE) \citep{Sudret2021}. For a given coordinate vector $\mathbf{x}$, we seek a PCE of the form:
\begin{equation}
f\left(\mathbf{x}, \tilde{\bm{\theta}}\right) \approx \sum_{\bm{\alpha} \in \mathcal{A}} c_{\bm{\alpha}}(\mathbf{x}) \, \Psi_{\bm{\alpha}}\left(\tilde{\bm{\theta}}\right),
\label{eq:pce_surr}
\end{equation}
where $\{ \Psi_{\bm{\alpha}}(\tilde{\bm{\theta}}) \} $ is a multivariate orthogonal polynomial basis with respect to the probability measure of $ \tilde{\bm{\theta}}$, $\bm{\alpha} \in \mathbb{N}_0^n$ is a multi-index defining the total degree of each multivariate polynomial, $\mathcal{A} \subset \mathbb{N}_0^n$ is the finite set of indices used in the expansion (e.g., corresponding to polynomials up to total degree $p$), $c_{\bm{\alpha}}(\mathbf{x}) \in \mathbb{R}$ are the PCE coefficients, which are functions of the input coordinate $\mathbf{x}$. The polynomials $\Psi_\alpha$ will be chosen to be orthonormal following Askey's scheme based on the input distributions of $\tilde{\bm{\theta}}$ \citep{Xiu2002} such that $\left\langle \Psi_{\bm{\alpha}}, \Psi_{\bm{\alpha}} \right\rangle =1$ with $\left\langle \Psi_{\bm{\alpha}}, \Psi_{\bm{\alpha}} \right\rangle = \int_{\mathbb{R}^n} \Psi_{\bm{\alpha}}^2(\bm{\theta}) \, \pi_{\tilde{\bm{\theta}}}(\bm{\theta}) \, d\bm{\theta}$.

To approximate  $c_{\bm{\alpha}}(\mathbf{x})$ , we use a pseudo-spectral projection approach based on Gaussian quadrature with $Q$ quadrature points, such that the coefficients of the PCE are obtained as
\begin{equation}
	c_{\bm{\alpha}}(\mathbf{x}) = \sum_{q=1}^{Q} w^{(k)} f\left(\mathbf{x}, \bm{\theta}^{(q)}\right) \Psi_{\bm{\alpha}}\left(\bm{\theta}^{(q)}\right), 
\end{equation}
where $\left\{ \bm{\theta}^{(q)}, w^{(q)} \right\}_{q=1}^{Q}$ are the quadrature nodes and weights associated with $\pi_{\tilde{\bm{\theta}}}$. For a collection of input points  $X = \left\{ \mathbf{x}^{(1)}, \dots, \mathbf{x}^{(N)} \right\}$, the model is evaluated at each  $\mathbf{x}^{(j)}$  for all quadrature nodes  $\bm{\theta}^{(q)}$. A single model simulation yields
\begin{equation}
f\left(X, \bm{\theta}^{(q)}\right) = 
\begin{bmatrix}
f\left(\mathbf{x}^{(1)}, \bm{\theta}^{(q)}\right) \\
f\left(\mathbf{x}^{(2)}, \bm{\theta}^{(q)}\right) \\
\vdots \\
f\left(\mathbf{x}^{(N)}, \bm{\theta}^{(q)}\right)
\end{bmatrix} \in \mathbb{R}^N,
\end{equation}
which allows constructing all  $c_{\bm{\alpha}}\left(\mathbf{x}^{(j)}\right)$ simultaneously for only $Q$ evaluations of the model. As full model evaluations $f$ are typically computationally expensive, evaluations of the mean $\mathbf{m}_\mathrm{GP}$ of the surrogate model $\tilde{f}$ will be used instead. The PCE provides a full surrogate model for the stochastic response $\mathbf{m}_\mathrm{GP}\left(\tilde{\bm{\theta}}\right)=\mathbb{E}\left[\tilde{f}\left(X, \tilde{\bm{\theta}}\right)\right]$ across all input locations $X = \{\mathbf{x}^{(1)}, \dots, \mathbf{x}^{(N)}\}$.

The approximation is expressed as
\begin{equation}
\mathbf{m}_\mathrm{GP}\left(\tilde{\bm{\theta}}\right) \approx \sum_{\bm{\alpha} \in \mathcal{A}} \mathbf{c}_{\bm{\alpha}} \, \Psi_{\bm{\alpha}}\left(\tilde{\bm{\theta}}\right),
\end{equation}
where each $\mathbf{c}_{\bm{\alpha}} \in \mathbb{R}^N$  is a vector of PCE coefficients evaluated at all input points such as
\begin{equation}
\mathbf{c}_{\bm{\alpha}} = 
\begin{bmatrix}
c_{\bm{\alpha}}\left(\mathbf{x}^{(1)}\right) \\
c_{\bm{\alpha}}\left(\mathbf{x}^{(2)}\right) \\
\vdots \\
c_{\bm{\alpha}}\left(\mathbf{x}^{(N)}\right)
\end{bmatrix}.
\end{equation}

From this expansion, the mean at the observations $\mathbf{m}_{\mathrm{PCE}}(\bm{\theta}) \in \mathbb{R}^N$ and the variances at observations  $\bm{\sigma}_{\mathrm{PCE}}^2 \in \mathbb{R}^N$  of the model response over $X$ are obtained directly as components of the surrogate:
\begin{align}
    \mathbf{m}_{\mathrm{PCE}}(\bm{\theta}_\mathrm{ext})&=\mathbb{E}\left[\mathbf{m}_\mathrm{GP}\left(\tilde{\bm{\theta}}\right)\right]\approx\mathbf{c}_{\bm{0}},\\
    \bm{\sigma}_{\mathrm{PCE}}^2(\bm{\theta}_\mathrm{ext})&=\mathbb{V}\left[\mathbf{m}_\mathrm{GP}\left(\tilde{\bm{\theta}}\right)\right]\approx\sum_{\bm{\alpha} \neq \bm{0}} \mathbf{c}_{\bm{\alpha}}^2 \, \left\langle \Psi_{\bm{\alpha}}, \Psi_{\bm{\alpha}} \right\rangle,\label{eq:pce_var}
\end{align}
where the square $\mathbf{c}_{\bm{\alpha}}^2$ is computed component-wise, and therefore the covariance matrix is $\Sigma_{\mathrm{PCE}}=\bm{\sigma}_{\mathrm{PCE}}^2(\bm{\theta}_\mathrm{ext})I$, assuming independent outputs. The last expressions follow immediately from the orthogonality of the polynomial basis. Further statistical moments can be extracted, but, once the PCE is constructed, it is possible to approximate the full statistical behaviour of $f\left(X, \tilde{\bm{\theta}}\right)$, without requiring any additional sampling or postprocessing.

Since the response of $\tilde{f}\left(X,\tilde{\bm{\theta}}\right)$ is stochastic and depends on the extrinsic parameters $\bm{\theta}_{\mathrm{ext}}$, one can identify those parameters for which the induced variance of $f\left(X,\tilde{\bm{\theta}}\right)$ provides a quantitative measure of MFU. To do so, we adapt the likelihood term defined in Equation \ref{eq:likelihood_gaussian} to include the variance of the prediction under the assumption that the response follows a normal distribution. The resulting likelihood is
\begin{equation}
    \mathcal{L}_{\mathrm{IN}}(\bm{\theta}_\mathrm{ext})=(2\pi)^{-\frac{N}{2}}\det\left(\Sigma_{\mathrm{PCE}}\right)^{-\frac{1}{2}}\exp\left(-\frac{1}{2}\left\|\mathbf{y}-\mathbf{m}_{\mathrm{PCE}}(\bm{\theta}_\mathrm{ext})\right\|^2_{\Sigma_{\mathrm{PCE}}^{-1}}\right).
    \label{eq:embedded_likelihood}
\end{equation}

In a given iteration of an MCMC loop, samples of $\bm{\theta}_\mathrm{ext}$ are drawn, then $\tilde{\bm{\theta}}$ is defined based on them, the PCE for $\tilde{f}\left(X,\tilde{\bm{\theta}}\right)$ is constructed using evaluations of the forward model at a set of quadrature points, the statistical moments of the PCE  $\mathbf{m}_{\mathrm{PCE}}$ and $\Sigma_{\mathrm{PCE}}$ are computed, and finally the likelihood $\mathcal{L}_{\mathrm{IN}}(\bm{\theta}_\mathrm{ext})$ is evaluated to obtain a sample of the posterior distribution $\pi\left(\bm{\theta}_\mathrm{ext}|\mathbf{y}\right)$.

\subsection{Influence analysis}
\label{sec2.5}
\subsubsection{$\phi$-divergence with full posterior}
Let $Y=(y_1,\dots,y_n)$ denote the whole dataset of observations. Let $S$ denote a nonempty proper subset of indices $\{1,\dots,n\}$ and let its complement be $S^c=\{1,\dots,n\}\setminus S$. We denote by $Y_S=\{Y_i:i\in S\}$ and $Y_{S^c}=\{Y_i:i\in S^c\}$ the corresponding subvectors of observations. We investigate the effect of removing $S$ from the data set on the posterior distribution of the model parameters. Assuming that $\pi(\bm{\theta}\mid Y)$ has global support, we define the posterior distributions
\begin{equation}
\pi(\bm{\theta} \mid Y_S) \propto \pi(\bm{\theta})\pi(Y_S \mid \bm{\theta})
\quad \text{and} \quad
\pi(\bm{\theta} \mid Y_{S^c}) \propto \pi(\bm{\theta})\pi(Y_{S^c} \mid \bm{\theta}).
\end{equation}

The \textit{influence} of $S$ can be quantified using a $\phi$-divergence \citep{Weiss1996a}. The $\phi$-divergence between the two probability density functions $\pi(\bm{\theta} \mid Y)$ and $\pi(\bm{\theta} \mid Y_{S^c})$ is defined as
\begin{equation}
D_\phi(S) =\int\phi\left(\frac{\pi(\bm{\theta} \mid Y_{S^c})}{\pi(\bm{\theta} \mid Y)}\right)\pi(\bm{\theta} \mid Y)\, d\bm{\theta},
\end{equation}
where $\phi: \mathbb{R}_0^+ \to \mathbb{R}$ is a convex function satisfying $\phi(1)=0$. A common choice for $\phi$ is the reverse Kullback-Leibler (KL) divergence, obtained by setting $\phi(\cdot) = -\log(\cdot)$. For simplification, we will denote by $D_\phi(S)$ the $\phi$-divergence of the posterior distribution obtained with the full data $Y$ and after excluding the subset $S$, using the reverse KL-divergence.

This formulation has the advantage that it does not require recomputing the posterior distribution for each subset $Y_{S^c}$. Instead, it can be evaluated directly from posterior samples and their likelihood values \citep{Weiss1996a, Zhu2012}. The data vector $Y=(Y_S,Y_{S^c})$ fulfills the chain rule of probability
\begin{equation}
\pi(Y\mid\bm{\theta})=\pi(Y_S \mid Y_{S^c}, \bm{\theta})\,\pi(Y_{S^c} \mid \bm{\theta}).
\end{equation}
Therefore, we define the ratio of likelihoods between using all data $Y$ and using only $Y_{S^c}$ as
\begin{equation}
\pi_S(\bm{\theta}) =\frac{\pi(Y \mid \bm{\theta})}{\pi(Y_{S^c} \mid \bm{\theta})}
=\pi(Y_S \mid Y_{S^c}, \bm{\theta}).
\label{eq:influence_ratio}
\end{equation}

Then, applying Bayes' law and Equation \ref{eq:influence_ratio},
\begin{eqnarray}
\pi(\bm{\theta} \mid Y_{S^c})
&=& \frac{\pi(Y_{S^c} \mid \bm{\theta})\pi(\bm{\theta})}{\int\pi(Y_{S^c} \mid \bm{\theta})\pi(\bm{\theta})d\bm{\theta}}\\&=&\frac{\pi(Y \mid \bm{\theta})\pi(\bm{\theta})[\pi_S(\bm{\theta})]^{-1}}
{\int \pi(Y \mid \bm{\theta})\pi(\bm{\theta})[\pi_S(\bm{\theta})]^{-1} d\bm{\theta}}\\&=&\frac{\pi(\bm{\theta}\mid Y)\pi(Y)[\pi_S(\bm{\theta})]^{-1}}
{\int \pi(\bm{\theta}\mid Y)\pi(Y)[\pi_S(\bm{\theta})]^{-1} d\bm{\theta}}\\&=&\frac{\pi(\bm{\theta}\mid Y)[\pi_S(\bm{\theta})]^{-1}}
{\int \pi(\bm{\theta}\mid Y)[\pi_S(\bm{\theta})]^{-1} d\bm{\theta}}\\&=&\frac{\pi(\bm{\theta}\mid Y)[\pi_S(\bm{\theta})]^{-1}}
{\mathbb{E}_{\bm{\theta} \mid Y}\left[\pi_S(\bm{\theta})^{-1}\right]},
\end{eqnarray}
where $\mathbb{E}_{\bm{\theta} \mid Y}$ denotes expectation with respect to the posterior distribution $\pi(\bm{\theta} \mid Y)$. Substituting this expression into the definition of $D_\phi(S)$ and simplifying yields the following expression for the influence of $S$
\begin{equation}
D_\phi(S)=\int \phi\left(\frac{[\pi_S(\bm{\theta})]^{-1}}{\mathbb{E}_{\bm{\theta} \mid Y}\left[\pi_S(\bm{\theta})^{-1}\right]}\right) \pi(\bm{\theta} \mid Y) d\bm{\theta}= \mathbb{E}_{\bm{\theta} \mid Y}\left[\phi\left(\frac{[\pi_S(\bm{\theta})]^{-1}}{\mathbb{E}_{\bm{\theta} \mid Y}\left[\pi_S(\bm{\theta})^{-1}\right]}\right)\right].
\end{equation}
Finally, substituting the influence function $\phi(u)=-\log(u)$ we get
\begin{equation}
D_\phi(S)=\log \mathbb{E}_{\bm{\theta} \mid Y}\left[\pi_S(\bm{\theta})^{-1}\right]
+ \mathbb{E}_{\bm{\theta} \mid Y}\left[\log \pi_S(\bm{\theta})\right].
\label{eq:influence_global}
\end{equation}
This influence measure can be computed for different sets $S$ to evaluate the relative influence of some observations over others on the inferred posterior. See Appendix \ref{ap:influence_global} for a stable estimation of the global influence from posterior samples.

\subsubsection{$\phi$-divergence on the marginal posterior using Kernel Density Estimation}
We are interested in extending the formulation of $D_\phi$ of Equation \ref{eq:influence_global} for a single parameter component $\theta_j$. Since $\pi_S(\bm{\theta})$ is not a probability density, marginalization must be defined with respect to the joint posterior distribution. We therefore consider the conditional posterior expectation of $\pi_S(\bm{\theta})$ given $\theta_j$. The posterior expectation conditional on the value of $\theta_j$ is
\begin{equation}
\pi_S^{(j)}(\theta_j) = \mathbb{E}_{\bm{\theta}\mid Y}\left[\pi_S(\bm{\theta}) \mid \theta_j\right]= \int \pi_S(\theta_j,\theta_{-j})  p(\theta_{-j}\mid\theta_j, Y) d\theta_{-j},
\end{equation}
where $\theta_{-j}$ denotes all components except $\theta_j$. An alternative would be to marginalize the perturbed $\pi(Y_{S^c} \mid \bm{\theta})$ and reference posteriors $\pi(Y \mid \bm{\theta})$ with respect to $\theta_{-j}$ and compute a divergence directly between the resulting marginal distributions of $\theta_{j}$. 

In practice, we approximate $\pi_S^{(j)}(\theta_j)$ from a posterior sample $\left\lbrace\bm{\theta}^{(i)}\right\rbrace_{i=1}^N$ using a one-dimensional weighted kernel density in the $\theta_j$ coordinate \citep{Hastie2009}. Let $K:\mathbb{R}\to\mathbb{R}^+$ be a kernel and $h>0$ the bandwidth. We define the Gaussian kernel $K_h(u)=\frac{1}{h}K\left(\frac{u}{h}\right)$ with $K(u)=(2\pi)^{-1/2}e^{-u^2/2}$. Using sample values $\theta_j^{(i)}$ and weights $w_i\propto\pi_S\left(\bm{\theta}^{(i)}\right)$ computed from full-sample likelihoods, we compute the Nadaraya-Watson estimator \citep{Nadaraya1964,Watson1964} of the conditional expectation as
\begin{equation}
\widehat{\pi}_S^{(j)}(\theta_j)
= \frac{\sum\limits_{i=1}^N w_iK_h\left(\theta_j-\theta_j^{(i)}\right)}
{\sum\limits_{i=1}^N K_h\left(\theta_j-\theta_j^{(i)}\right)},
\end{equation}
where the bandwith $h$ is chosen following Scott's rule \citep{Scott2008}. We then insert this smoothed conditional ratio into the $\phi$-divergence influence functional. The marginal influence for component $j$ is
\begin{equation}
D_\phi^{(j)}(S)
= \log \mathbb{E}_{\theta_j\mid Y}\left[\widehat{\pi}_S^{(j)}(\theta_j)^{-1}\right]
+\mathbb{E}_{\theta_j\mid Y}\left[\log \widehat{\pi}_S^{(j)}(\theta_j)\right].
\label{eq:influence_kde}
\end{equation}
See Appendix \ref{ap:influence_kde} for a stable estimation of the marginal influence from posterior samples using the KDE approach.

\subsubsection{$\phi$-divergence on the marginal posterior using fixed-mean parameter estimators}
It is also of interest to compute the marginal influence on one parameter when the others are fixed to their estimated values. In this case, this marginal influence will be approximated by fixing them to the mean of their posterior samples. We first compute such mean of $\theta_{-j}$ as
\begin{equation}
\bar{\theta}_{-j} = \mathbb{E}_{\bm{\theta}\mid Y}\left[\theta_{-j}\right]
\approx \frac{1}{N}\sum_{i=1}^N \theta_{-j}^{(i)}.
\end{equation}
The fixed-mean perturbation as a function of $\theta_j$ is the pointwise evaluation
\begin{equation}
\pi_S^{(\text{fix})}(\theta_j) = \pi_S(\theta_j,\bar{\theta}_{-j})
= \frac{\pi(Y\mid \theta_j,\bar{\theta}_{-j})}{\pi(Y_{S^c}\mid \theta_j,\bar{\theta}_{-j})}.
\end{equation}
Using this plug-in perturbation, we define the fixed-mean influence
\begin{equation}
D_\phi^{(j),(\text{fix})}(S)
= \log \mathbb{E}_{\theta_j\mid Y}\left[\pi_S^{(\text{fix})}(\theta_j)^{-1}\right]
+ \mathbb{E}_{\theta_j\mid Y}\left[\log \pi_S^{(\text{fix})}(\theta_j)\right].
\label{eq:influence_fixed}
\end{equation}
See Appendix \ref{ap:influence_fixed_mean} for a stable estimation of the marginal influence from posterior samples using the fixed mean approach.

\subsection{Uncertainty propagation, separability and identifiability}
\label{sec2.6}

Unlike classical approaches, the embedded formulation enables the propagation of quantified model‐form uncertainty through an extended parameter set. Let $g$ be a real‐valued map, analogous to $f$, defined as
\begin{equation}
    g:\mathcal{X}_g\times\Theta_g\times\Lambda_g \subset \mathbb{R}^q \to \mathbb{R}^m,
    \qquad
    (\mathbf{x},\bm{\theta},\bm{\lambda}) \mapsto g(\mathbf{x},\bm{\theta},\bm{\lambda}),
\end{equation}
where $\mathbf{x}\in\mathcal{X}_g$ denotes the input coordinates at which the model response is evaluated, $\bm{\theta}\in\Theta_g\subseteq\Theta$ is the vector of parameters previously calibrated using model $f$, and $\bm{\lambda}\in\Lambda_g$ is a vector of additional parameters that are not identified in the calibration stage. In contrast to $\bm{\theta}$, the parameters $\bm{\lambda}$ are quantities of direct interest that must be inferred from measurement data in the physical system, for example damage‐related state variables.

The model $g$ represents a downstream prediction task in which the calibrated parameters $\bm{\theta}$ are reused under a different modeling context. When the parameters $\bm{\theta}$ are calibrated with embedded uncertainties, they are represented by random variables $\tilde{\bm{\theta}}$, rendering the model response $g(\mathbf{x},\tilde{\bm{\theta}},\bm{\lambda})$ inherently stochastic. For each fixed value of $\bm{\lambda}$, the uncertainty in $\tilde{\bm{\theta}}$ induces a predictive distribution of the model output. The central objective is to assess how this propagated uncertainty affects the ability to distinguish between different candidate values of $\bm{\lambda}$ based on the resulting predictive distributions.

In this work, the model $g$ is a finite element model of a reinforced concrete beam used to investigate whether measurements from a SHM system can be linked to corresponding values of the model‐state parameters $\bm{\lambda}$. Specifically, $\bm{\lambda}$ encodes the location of a tendon failure, parameterized through the depth of the tendon break. The predictive distributions of the structural response for different values of $\bm{\lambda}$ are compared against observed strain patterns to evaluate how accurately the damaged tendon can be identified in the presence of embedded model‐form uncertainty. As in the calibration stage, a PCE surrogate is employed to approximate the stochastic model response, following Equations~\ref{eq:pce_surr}–\ref{eq:pce_var}. This surrogate enables efficient propagation of the uncertainty in $\tilde{\bm{\theta}}$ for each candidate value of $\bm{\lambda}$, yielding the predictive distributions required for subsequent identifiability and influence analyses.

\begin{figure}[htbp]
    \centering
    \includegraphics[width=\linewidth]{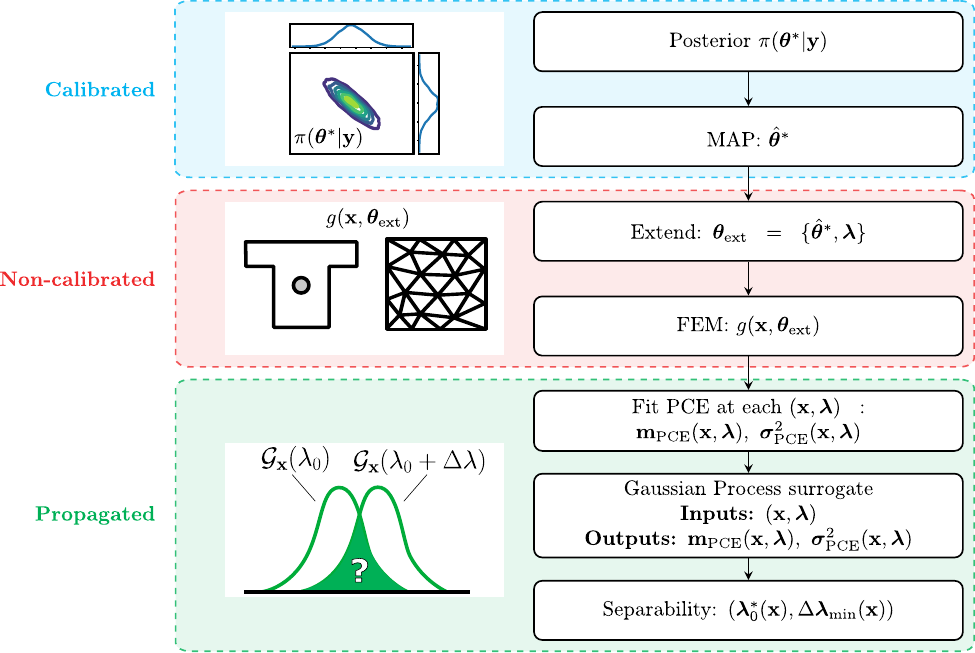}
    \caption{Flow diagram of uncertainty propagation and identifiability analysis. 
    The calibrated parameters $\bm{\theta}^*$ are extended with a set of non-calibrated parameters $\bm{\lambda}$, propagated through the FEM model $g$, approximated via PCEs, and used to train GP surrogates for separability analysis.}
    \label{fig:separability_flow}
\end{figure}

Using the calibrated parameter set $\bm{\theta}$, we simulate system responses $g(\mathbf{x}, \tilde{\bm{\theta}}, \bm{\lambda})$ for various hypothesized values of $\bm{\lambda}$. Running these simulations over a range of possible breakage locations generates a family of predictive distributions, each representing the expected system response for a particular damage scenario. These predictive distributions provide the basis for defining detection thresholds within the monitoring framework, which can be tailored to specific spatial locations $\mathbf{x}$ in the structure. A diagram of the proposed workflow is represented in Figure \ref{fig:separability_flow}.

However, the MFU embedded in $\tilde{\bm{\theta}}$ leads to overlapping output distributions for different values of $\bm{\lambda}$. When the predictive distributions for $g\left(\mathbf{x}, \tilde{\bm{\theta}}, \bm{\lambda}_1\right)$ and $g\left(\mathbf{x}, \tilde{\bm{\theta}}, \bm{\lambda}_2\right)$, corresponding to $\bm{\lambda}_1$ and $\bm{\lambda}_2 = \bm{\lambda}_1 + \Delta\bm{\lambda}$, are not statistically distinguishable, it becomes impossible to unambiguously infer the change in $\bm{\lambda}$ from observed data. This is the essence of the identifiability problem. Therefore, a central objective is to quantify the minimal detectable perturbation $\Delta\bm{\lambda}_{\min}$ such that the induced change in system response exceeds the uncertainty envelope defined by the calibrated model. This enables a principled definition of detection thresholds that are both sensitive and robust, grounded in the probabilistic structure of the FEM response. The identifiability of a change in the parameter vector $\bm{\lambda}$ depends not only on the magnitude of the perturbation $\Delta\bm{\lambda}$, but also on the current operating point $\bm{\lambda}_0$ and the direction of change (i.e., whether $\bm{\lambda}_0 \pm \Delta\bm{\lambda}$). Due to the nonlinear and probabilistic nature of the forward model $g(\mathbf{x}, \tilde{\bm{\theta}}, \bm{\lambda})$, the propagated response distributions may differ significantly depending on both $\bm{\lambda}_0$ and the sign of the perturbation.

The identifiability analysis is conducted locally at each point $\mathbf{x}$ in the spatial domain. For a given $\mathbf{x}$, the objective is to determine the smallest perturbation $\Delta\bm{\lambda}$ that leads to a statistically distinguishable change in the model response, as well as the corresponding parameter configuration $\bm{\lambda}_0^\ast$ for which this minimal change is most difficult to detect. Let the response distribution at $\mathbf{x}$ be defined as $\mathcal{G}_{\mathbf{x}}(\bm{\lambda}) =  g(\mathbf{x}, \tilde{\bm{\theta}}, \bm{\lambda})$. The chosen identifiability criterion is based on the principle of non-overlap of the confidence intervals at 95\%. We define the spatially-localized maximin optimization problem as
\begin{equation}
(\bm{\lambda}_0^\ast(\mathbf{x}), \Delta\bm{\lambda}_{\min}(\mathbf{x})) = \arg \max_{\bm{\lambda}_0 \in \Lambda_g} \; \min_{\Delta\bm{\lambda}>0} 
\|\Delta\bm{\lambda}\| ~~\text{subject to}~~
\left\{
\begin{aligned}
&\text{CI}_{0.95}[\mathcal{G}_{\mathbf{x}}(\bm{\lambda}_0)] \cap \text{CI}_{0.95}[\mathcal{G}_{\mathbf{x}}(\bm{\lambda}_0 + \Delta\bm{\lambda})] = \emptyset, \\
&\text{CI}_{0.95}[\mathcal{G}_{\mathbf{x}}(\bm{\lambda}_0)] \cap \text{CI}_{0.95}[\mathcal{G}_{\mathbf{x}}(\bm{\lambda}_0 - \Delta\bm{\lambda})] = \emptyset.
\end{aligned}
\right.
\label{eq:maximin_problem}
\end{equation}

Here, $\Delta\bm{\lambda}_{\min}(\mathbf{x})$ is the smallest detectable parameter shift at point $\mathbf{x}$ and $\bm{\lambda}_0^\ast(\mathbf{x})$ is the least favorable configuration, i.e., the point in parameter space where distinguishability is hardest under the confidence interval criterion. This formulation ensures that detection thresholds can be defined locally and conservatively, based on the worst-case identifiability scenario at each spatial location. In practical terms, solving this problem enables spatially varying sensitivity maps to be constructed, which guide sensor placement and interpretation of system responses under monitoring conditions.

Due to the high computational cost of evaluating the FEM model $g(\mathbf{x}, \bm{\theta}, \bm{\lambda})$, a surrogate modeling strategy is adopted to make the identifiability optimization tractable. The surrogate captures the dependence of the distribution of $g$ on the parameters $\bm{\lambda}$, accounting for the uncertainty introduced by the parameters $\tilde{\bm{\theta}}$. For each pair $(\mathbf{x}, \bm{\lambda})$, the distribution $\mathcal{G}_{\mathbf{x}}(\bm{\lambda}) =g\left(\mathbf{x}, \tilde{\bm{\theta}}, \bm{\lambda}\right)$ is approximated using a PCE with $\tilde{\bm{\theta}}$ as input, yielding the conditional mean $\mathbf{m}_\mathrm{PCE}(\mathbf{x}, \bm{\lambda})$ and variance $\bm{\sigma}_\mathrm{PCE}^2(\mathbf{x}, \bm{\lambda})$ of the model response as formulated in Equations \ref{eq:pce_surr} to \ref{eq:pce_var}. These statistics are then modeled using two independent GP regressors, one for the mean and one for the variance, whose implementation details will be reviewed in Section \ref{sec2.3}. The training dataset for the GP surrogates is constructed as follows: a set of spatial locations $X=\left[\mathbf{x}_i\right]_{i=1}^{N_x}$ is selected as a subset of the mesh nodes used in the FEM simulation. Then, a discrete grid $\{\bm{\lambda}_j\}_{j=1}^{N_\lambda}\subset\Lambda_g$ is chosen to cover the admissible domain $\Lambda_g$ of the uncertain parameters. For each training pair $(\mathbf{x}_i, \bm{\lambda}_j)$, the corresponding PCE is constructed to compute the mean and variance of $\mathcal{G}_{\mathbf{x}}(\bm{\lambda})$, which are then used to train the GPs. This dataset can be formulated as
\begin{equation}
\mathcal{D}_{\text{train}} = \left\{\left. (X, \bm{\lambda}_i), \; \left(\mathbf{m}_\mathrm{PCE}(X,\bm{\lambda}_i), \bm{\sigma}^2_\mathrm{PCE}(X,\bm{\lambda}_i)\right) \right| i=1,...,N_\lambda\right\}.
\end{equation}
Only $N_\lambda\cdot N_Q$ evaluations of $g$ are required, where $N_Q$ is the number of quadrature points for the PCE. Once trained, the GPs provide efficient predictions $\mathbf{m}_\mathrm{PCE}(\mathbf{x}, \bm{\lambda})$ and $\bm{\sigma}_\mathrm{PCE}^2(\mathbf{x}, \bm{\lambda})$ at any $(\mathbf{x}, \bm{\lambda})$ pair within the domain of interest. The 95\% confidence intervals for the response distributions are then approximated as
\begin{equation}
\text{CI}_{0.95}[\mathcal{G}_{\mathbf{x}}(\bm{\lambda})] \approx \left[ \tilde{\mathbf{m}}_\mathrm{PCE}(\mathbf{x}, \bm{\lambda}) - 1.96 \, \tilde{\bm{\sigma}}_\mathrm{PCE}^2(\mathbf{x}, \bm{\lambda}), \; \tilde{\mathbf{m}}_\mathrm{PCE}(\mathbf{x}, \bm{\lambda}) + 1.96 \, \tilde{\bm{\sigma}}_\mathrm{PCE}^2(\mathbf{x}, \bm{\lambda}) \right],
\label{eq:ci_with_gp}
\end{equation}
where $\tilde{\mathbf{m}}_\mathrm{PCE}$ and $\tilde{\bm{\sigma}}_\mathrm{PCE}^2$ are the statistical moments estimated by their respective GPs. The variance introduced by the GP is disregarded for simplicity, as $\mathcal{D}_\mathrm{train}$ will be sufficiently dense. The confidence interval approximation of Equation \ref{eq:ci_with_gp} can be directly plugged into the optimization problem of Equation \ref{eq:maximin_problem}. The domain $\Lambda_g$ is discretized as well for the maximization part of the optimization problem, choosing the same fixed grid $\{\bm{\lambda}_j\}_{j=1}^{N_\lambda}$ as used in the GP training, improving the efficiency of the evaluations. The minimization problem for $\Delta\bm{\lambda}$ is solved using the Nelder-Mead algorithm, while the maximization of $\bm{\lambda}_0$ is performed by comparison in the grid. 
This formulation enables efficient computation of the minimal detectable changes across the spatial domain, making it feasible to generate full-field maps of identifiability under uncertainty.

If separability is not achieved within a predefined maximum perturbation $\Delta\lambda_{\max}$, the optimization problem in Equation \ref{eq:maximin_problem} is infeasible. Even in that case, we can extract some insight from partial solutions. To address this, we first check separability at $\lambda\pm \Delta\lambda_{\max}$ for each candidate $\lambda$. If separability is already satisfied at this level, then the optimization proceeds as usual. Otherwise, we fix $\Delta\lambda = \Delta\lambda_{\max}$ and quantify the degree of indistinguishability by computing the overlap between the two distributions. The overlap integral is computed to measure the amount of shared probability mass between the two distributions corresponding to $\lambda$ and $\lambda\pm\Delta\lambda_{\max}$ as
\begin{equation}
\begin{aligned}
O(\mathbf{x},\lambda;\Delta\lambda_{\max})
\equiv&
\frac{1}{2}\left(
\int_{-\infty}^{\infty}
\min\Big\{ \,
\mathcal{G}_{\mathbf{x}}(\bm{\lambda}), \;
\mathcal{G}_{\mathbf{x}}(\bm{\lambda+\Delta\lambda_{\max}}) \,
\Big\}\, \mathrm{d}y\right.\\
&\left.+\int_{-\infty}^{\infty}
\min\Big\{ \,
\mathcal{G}_{\mathbf{x}}(\bm{\lambda}), \;
\mathcal{G}_{\mathbf{x}}(\bm{\lambda-\Delta\lambda_{\max}}) \,
\Big\}\, \mathrm{d}y\right).
\end{aligned}
\label{eq:overlap_integral}
\end{equation} 
where $y$ represents the output of the (i.e. the support of $\mathcal{G}_{\mathbf{x}}$) over which the probability mass is defined. In practice, $\mathcal{G}_{\mathbf{x}}$ will be modelled as a normal random variable defined by the PCE, and the integrals will be calculated by numerical integration in the range of $\tilde{\mathbf{m}}_\mathrm{PCE}\pm4\tilde{\bm{\sigma}}_\mathrm{PCE}^2$, truncating the distribution everywhere else.

From the collection of overlaps across the parameter grid, summary statistics can be extracted as
\begin{equation}
\begin{aligned}
O_{\min}(\mathbf{x}) &= \min_{\lambda\in\Lambda_g} O(\mathbf{x},\lambda;\Delta\lambda_{\max}),\\
O_{\max}(\mathbf{x}) &= \max_{\lambda\in\Lambda_g} O(\mathbf{x},\lambda;\Delta\lambda_{\max}),\\
R_O(\mathbf{x}) &= O_{\max}(\mathbf{x}) - O_{\min}(\mathbf{x}).
\end{aligned}
\label{eq:overlap_stats}
\end{equation}
These values provide complementary information: $O_{\min}$ indicates the best-case separability, $O_{\max}$ the worst-case indistinguishability, and $R_O$ quantifies the variability of overlap across $\Lambda_g$. The complete procedure is summarized in Algorithm \ref{alg:delta_overlap}.

\begin{algorithm}[t]
\caption{Minimal detectable change with separability check at $\Delta\lambda_{\max}$}
\label{alg:delta_overlap}
\begin{algorithmic}[1]
\REQUIRE Spatial nodes $X=\{\mathbf{x}_i\}_{i=1}^{N_x}$, parameter grid $\Lambda_g=\{\lambda_j\}_{j=1}^{N_\lambda}$, 
GP surrogates for $\tilde{\mathbf m}_\mathrm{PCE},\tilde{\bm\sigma}_\mathrm{PCE}^2$, 
maximum perturbation $\Delta\lambda_{\max}$, optimizer $\mathcal{O}$.
\ENSURE For each $\mathbf{x}_i$: $(\bm{\lambda}_0^\ast(\mathbf{x}_i), \Delta\bm{\lambda}_{\min}(\mathbf{x}_i))$, overlap map $O(\mathbf{x}_i,\lambda_j)$, and $(O_{\min},O_{\max},R_O)$.
\vspace{2pt}
\FOR{each spatial node $\mathbf{x}_i \in X$}
  \FOR{each parameter $\lambda_j \in \Lambda_g$}
    \STATE \textbf{Check separability at $\Delta\lambda_{\max}$:}
      \begin{equation*}
      \text{Separable}\leftarrow\left\lbrace\begin{aligned}
&\text{CI}_{0.95}[\mathcal{G}_{\mathbf{x}_i}(\lambda_j)] \cap \text{CI}_{0.95}[\mathcal{G}_{\mathbf{x}_i}(\lambda_j + \Delta\lambda_{\max})] \stackrel{?}{=} \emptyset, \\
&\text{CI}_{0.95}[\mathcal{G}_{\mathbf{x}_i}(\lambda_j)] \cap \text{CI}_{0.95}[\mathcal{G}_{\mathbf{x}_i}(\lambda_j - \Delta\lambda_{\max})] \stackrel{?}{=} \emptyset.
\end{aligned}\right.
      \end{equation*}
  \ENDFOR
    \IF{$\text{Separable}~\forall\lambda_j\in\Lambda_g $} 
      \STATE \COMMENT{Perform optimization}
      \STATE Solve Eq.~\eqref{eq:maximin_problem} with optimizer $\mathcal{O}$ to obtain $(\bm{\lambda}_0^\ast(\mathbf{x}_i), \Delta\bm{\lambda}_{\min}(\mathbf{x}_i))$.
    \ELSE
      \STATE \COMMENT{Fix to $\Delta\lambda_{\max}$ and compute overlap}
      \STATE Set $\Delta\lambda^*(\mathbf{x}_i,\lambda_j) \leftarrow \Delta\lambda_{\max}$.
      \FOR{each parameter $\lambda_j \in \Lambda_g$}
        \STATE Compute overlap $O(\mathbf{x}_i,\lambda_j)$ using Eq.~\eqref{eq:overlap_integral}.
        \STATE Compute statistics $O_{\min},O_{\max},R_O$ with Eq.~\eqref{eq:overlap_stats}.
      \ENDFOR
    \ENDIF
\ENDFOR
\RETURN Maps of $\bm{\lambda}_0^\ast$, $\Delta\bm{\lambda}_{\min}$, $O$, $O_{\min},O_{\max},R_O$.
\end{algorithmic}
\end{algorithm}

\subsection{Gaussian Process Surrogate Model}
\label{sec2.3}
We consider the previously defined simulation $f$ defined at Equation~\ref{eq:def_forward_model} as the forward model to be calibrated that maps model parameters to observable system responses. Throughout this section, we construct a Gaussian Process (GP) surrogate $\tilde{f}$ to emulate the input–output behavior of the model.  
Let $\{\bm{\xi}_i \in \Omega \subset \mathbb{R}^d,~i=1,\dots,n\}$ denote a set of design points. Evaluations of the model at these points yield the training dataset
\begin{equation*}
    \mathcal{D} = \{ (\bm{\xi}_i, f(\bm{\xi}_i)) : i=1,\dots,n \}, \quad
    \Xi = 
    \begin{bmatrix}
    \bm{\xi}_1^\top & \cdots & \bm{\xi}_n^\top
    \end{bmatrix} \in \mathbb{R}^{d \times n}, \quad
    \mathbf{f} = 
    \begin{bmatrix}
    f(\bm{\xi}_1) & \cdots & f(\bm{\xi}_n)
    \end{bmatrix} \in \mathbb{R}^{m \times n}.
\end{equation*}
For each output component, a GP prior is placed as
\begin{equation}
	\tilde{f}(\bm{\xi})\sim\mathcal{GP}\left(\mu(\bm{\xi}),k(\bm{\xi},\bm{\xi'})\right)
\end{equation}
where $\mu:\Omega\to\mathbb{R}$ is the prior mean and $k:\Omega\times\Omega\to\mathbb{R}$ is a positive semi-definite covariance kernel \citep{Rasmussen2006}. Since the training outputs are standardized to zero mean for numerical stability, we take $m(\bm{\xi})\equiv0$.

Let $\Xi^*=\left[\bm{\xi}^*_1,...,\bm{\xi}^*_n\right]^\top$ be a set of of test points with corresponding outputs $\mathbf{f^*}=\tilde{f}(\Xi^*)$. The GP prior implies the joint distribution
\begin{equation}
    \begin{bmatrix}
    \mathbf{f}\\
    \mathbf{f^*}
    \end{bmatrix}\sim\mathcal{N}\left(\mathbf{0},\begin{bmatrix}
K(\Xi,\Xi) & K(\Xi,\Xi^*) \\
K(\Xi^*,\Xi) & K(\Xi^*,\Xi^*) \\
\end{bmatrix}\right)
\end{equation}
with $K(\Xi, \Xi') = \left[ k(\bm{\xi}_i, \bm{\xi}'_j) \right]$. Denoting the vector of the kernel evaluation for a given test point $\bm{\xi}^*$ with respect to the training points $\bm{\xi}_i$ as $\mathbf{k}(\bm{\xi}^*)=[k(\bm{\xi}^*,\bm{\xi}_1),\dots,k(\bm{\xi}^*,\bm{\xi}_n)]^\top$, the predictions for such point are
\begin{align}
    \mathbb{E}\left[\tilde{f}(\bm{\xi}^*)\right]&=\mathbf{k}(\bm{\xi}^*)^\top K(\Xi,\Xi)^{-1}\mathbf{f},\\
    \mathbb{V}\left[\tilde{f}(\bm{\xi}^*)\right]&=\mathbf{k}(\bm{\xi}^*,\bm{\xi}^*)-\mathbf{k}(\bm{\xi}^*)^\top K(\Xi,\Xi)^{-1}\mathbf{k}(\bm{\xi}^*).
\end{align}
The chosen kernel will be a radial basis function with added white noise:
\begin{equation}
    k(\xi_i, \xi_j) = \sigma_f^2 \exp\left(-\frac{\|\xi_i - \xi_j\|^2}{2\ell^2} \right) + \sigma_n^2 \delta_{ij},
\end{equation}
where $\sigma_f^2$ is the signal variance, $\ell$ is the correlation length-scale, $\sigma_n^2$ is the white noise variance and $\delta_{ij}$ is the Kronecker delta which is 1 if $i=j$ and 0 otherwise. Other kernels are possible but do not present any apparent advantages due to the expected smoothness of the target response surface.

Training a GP regressor as a surrogate model is reduced to solving the minimization problem 
\begin{equation}
    \bm{\theta}_{\mathrm{GP}}^* = \arg\min_{\bm{\theta}_{\mathrm{GP}}} \, -\log \pi(\mathbf{f} \mid \Xi, \bm{\theta}_{\mathrm{GP}}),
\end{equation}
where $\bm{\theta}_{\mathrm{GP}} = \{\sigma_f^2, \ell, \sigma_n^2\}$ denotes the set of hyperparameters of the kernel and $\pi(\mathbf{f} \mid \Xi, \bm{\theta}_{\mathrm{GP}})$ is the probability density function of the GP evaluated at the training outputs $\mathbf{f}$, which follows a multivariate normal distribution. This corresponds to maximizing the log marginal likelihood of the training data under the Gaussian process prior. The optimization is performed using gradient-based methods as implemented in scikit-learn's \texttt{GaussianProcessRegressor} \citep{scikit-learn2011}, which follows the standard algorithmic framework described in \cite{Rasmussen2006}. All models are trained using automatic hyperparameter tuning via marginal likelihood maximization with multiple restarts to avoid poor local optima.

Gaussian Processes constitute an appropriate surrogate modelling framework for the present application due to their sample efficiency and strong approximation capabilities for smooth, deterministic response surfaces. Their kernel-based structure enables accurate emulation of complex simulation outputs from a limited number of high-fidelity evaluations \citep{Forrester2008}. We employ GP surrogates in two distinct ways. The first approach approximates the response surfaces with respect to a set of parameters $\boldsymbol{\theta}\in\Theta\subset\mathbb{R}^d$ at a fixed set of spatial coordinates $X=\{\mathbf{x}_i\}_{i=1}^N$. The second approach constructs a single GP surrogate defined jointly over both coordinates and parameters.

In the first case, the input to the GP is restricted to the parameter vector, i.e.,
$\boldsymbol{\xi}=\{\boldsymbol{\theta}\}$. For each spatial coordinate $\mathbf{x}_i\in X$, a separate GP is trained and later evaluated as
\begin{equation}
	\tilde{f}(\mathbf{x}_i,\boldsymbol{\theta})	\sim \mathcal{GP}\!\left(m(\boldsymbol{\theta}), k(\boldsymbol{\theta},\boldsymbol{\theta}')\right),
\end{equation}
using the training dataset $\mathcal{D}=\{(\boldsymbol{\theta}_j,\, f(\mathbf{x}_i,\boldsymbol{\theta}_j))\}_{j=1}^n$. Since a single simulation produces $N$ observations (one per coordinate in $X$), only $n$ simulations are required to train the $N$ independent GPs, one for each fixed coordinate. The surrogate can therefore interpolate only in the parameter space $\Theta$, while predictions must be evaluated at the fixed coordinate set $X$. This formulation is suitable when all quantities of interest are measured or required at predefined observation locations. This first formulation will be used in the calibration of the parameters of the simulation model $f$ based on experiments. In general, including a GP in the calibration process will generate an additional uncertainty that should be considered in the likelihood. In some cases, it can be exploited for more efficient surrogate building \citep{Perrin2025}. However, for the current paper we will validate the GP showing that the variances in the validation dataset can be neglected, using it as a mean response surface.

In the second case, the GP input includes both the coordinate and parameter vectors, 
$\boldsymbol{\xi}=\{\mathbf{x},\boldsymbol{\theta}\}$, with  
$\boldsymbol{\xi}\in\Omega=\mathcal{X}\times\Theta\subset\mathbb{R}^d$, where $\mathcal{X}$ and $\Theta$ denote the coordinate and parameter spaces, respectively. A single GP is trained on this joint domain, enabling predictions across both spatial
coordinates and parameter values. This formulation is particularly suitable when predictions are needed on a dense or variable spatial grid, such as for computing full-field system responses. This second formulation will be used for the propagation of the uncertainties in the FEM model $g$ that will be used for additional parameters identification

\section{Application to a model for tendon breakage in reinforced concrete structures}
\label{sec3}
The methodology outlined in the former section establishes the foundation for an uncertainty-embedded calibration. This framework is applied to a tendon-breakage experiment in pretensioned concrete instrumented with distributed fiber-optic sensors (DFOS, see subsection~\ref{sec2.2.1}) on the surface, allowing the measurement of strain changes with high spatial resolution. This section provides a comprehensive description of the experimental setup and the accompanying finite element model. It also presents the resulting calibration, influence, and identifiability analysis. The objective of the identifiability assessment is to analyze how accurately can we distinguish the breakage of a tendon at different depths of a concrete specimen considering measurement noise and model uncertainty.

\subsection{Experimental Investigation}
\label{sec2.1} 
\subsubsection{Experimental Setup}
\begin{figure}[h]
\centering
\includegraphics[width=0.9\textwidth]{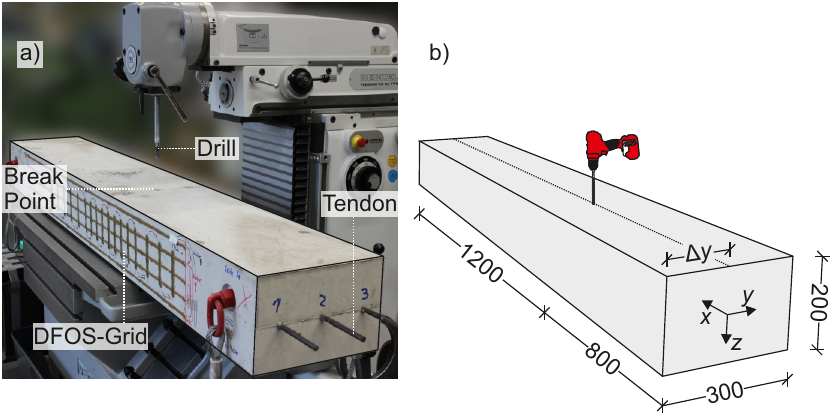}
\caption{Setup of the experimental investigation: specimen with drill and applied grid of distributed fiber optical sensors (a), sketch of the specimen with corresponding geometrical properties (b).}
\label{fig2}
\end{figure}

The experimental investigation was executed on a pretensioned concrete beam of rectangular cross section ($L \cdot b \cdot h=2000 \cdot 300 \cdot 200$\,[mm³]). In a stressing bed, a prestress $\sigma_{p,0} = 755$\,N/mm² was applied to three prestressing wires with smooth surfaces ($\varnothing_p=9.4$\,mm, St 1375/1570) and transferred to the beam after hardening of the concrete (28\,d). A normal strength concrete was employed. Based on three tested cubes, a mean compressive strength of $f_{cm,cube}=55.0$\,N/mm² was measured. 
The installation of a plastic tube between the designated breakage point of the tendon and the concrete surface ensured the accessibility of the tendon after pouring of the concrete. The location was initially shifted in the $x$-direction to ensure full re-anchoring on at least one site of the breakpoint. However, DFOS measurements conducted on the tendon during the experiment showed that complete re-anchorage is attained on both sides from the breakage \citep{Paul.2024b}. After application of the measurement equipment (details see subsection \ref{sec2.2.1}), the beam was transferred to the test rig, which is illustrated in Figure \ref{fig2}a. The breakage of the centric tendon (at $x=800$\,mm, $y=z=0$, cf. Figure \ref{fig2}b) was then induced using a drill. The complete separation was visually confirmed through the previously installed plastic tube and terminated the experimental investigation.

\subsubsection{Instrumentation}\label{sec2.2.1}
\begin{figure}[h]
\centering
\includegraphics[width=0.9\textwidth]{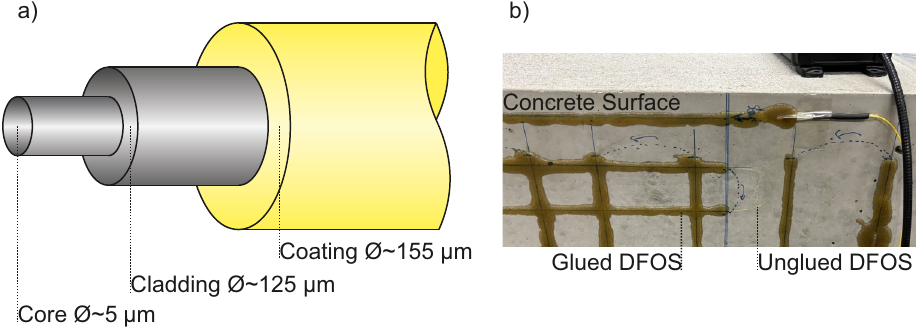}
\caption{Distributed fiber-optic sensor, as used in the instrumentation of the specimen: general structure of a sensor (a), sensor grid applied to the concrete surface (b).}
\label{fig3}
\end{figure}

DFOS \citep{Clau.2021, Speck.2019} were applied to the concrete surface \citep{Paul.2024b,Janiak.2023} to instrument the test specimen. The measurement principle of DFOS can be roughly described as follows: A light beam is emitted into the sensor core ($\varnothing \approx 5$\,µm; see Figure~\ref{fig3}a), which possesses a variable refractive index along the sensor due to micro-inclusions. This variation enables the certain identification of each measurement section through a distinctive "fingerprint". The backscatter is converted into the frequency domain by discrete Fourier transformation. Changes in strain $\Delta \varepsilon$ cause alterations of the wave length \citep{Samiec.2011, Barrias.2016}. 

A variety of sensor configurations is available. The main distinction between these sensors lies in their outer sensor layer (\textit{coating}, see Figure~\ref{fig3}a), which exerts a substantial influence on the transfer of strain from the sample to the sensor. Sensors with a polyimide, acrylate, or nylon coating are popular. Rigid coatings, such as polyimide (Young's modulus of $E_{\rm Polyimide} = 400E_{\rm Acrylate}$ \citep{Chapeleau.2021}), are recommended for high-precision measurements \citep{Herbers.2023}. The increased risk of breakage and sensor loss exhibited by stiff coatings in the presence of cracks can be disregarded here, as the induced damage to the prestressing steel usually does not result in concrete cracks \citep{Strater.2024}.

After the transfer of the prestressing force and removing the formwork, the DFOS is applied to the side surfaces of the concrete structure using a two-component epoxy-based adhesive as is shown in Figure~\ref{fig3}b. The application is executed parallel and orthogonal to the tendon axis as a 2D-grid, following a meandering pattern in which a single fiber traverses varying height levels. Loops are left unglued. The outer points of each layer of the grid are distinctly designated to their respective positions on the fiber strand according to the \textit{touch-to-locate method} \citep{Konertz.2019}. The interpolation of strains between measurement points and layers is achieved through bilinear interpolation \citep{Paul.2025}. The measurements were performed throughout the entire experiment with a measuring point spacing of 2.6\,mm and a frequency of 1\,Hz.

\subsection{Numerical Model}
\label{sec2.2}
A numerical model of the experimentally investigated specimen was developed by means of the finite element method (\textit{FEM}). For this purpose, the commercial software \textit{ABAQUS} is used. The concrete volume is modeled with cylindrical recesses corresponding to the positions of the tendons. Both the tendon and the concrete were discretized using three-dimensional solid elements with linear shape functions. A uniform element size is used in regions remote from the tendons, while progressive mesh refinement is applied towards the prestressing steel. This approach is consistent with established finite element modelling practices for reinforced and prestressed concrete structures \citep{vanMeirvenne.2018}. The analysis is performed using a static step-wise procedure, in which loads are applied incrementally under quasi-static conditions. Details are provided in the upcoming subsections.

\subsubsection{Constitutive Laws}\label{sec2.3.1}
The non-linear Concrete Damage Plasticity (CDP) constitutive law, which was developed by \citep{Lubliner.1989} and improved by \citep{Lee.1998} is employed for the concrete. It combines isotropic, non-associative plasticity with damage parameters to consider the degradation of the material. The effective stress $\overline{\boldsymbol{\sigma}}$ is calculated by means of the elasticity matrix $\boldsymbol{D}_0^{el}$, the total strain tensor $\boldsymbol{\varepsilon}$ and the plastic strain tensor $\boldsymbol{\varepsilon}^{pl}$
\begin{equation}\label{eq:EffStress}
    \overline{\boldsymbol{\sigma}} = \boldsymbol{D}^{el}_0 : \left( \boldsymbol{\varepsilon} - \boldsymbol{\varepsilon}^{pl} \right).
\end{equation}
For the computation of the Cauchy stress tensor $\boldsymbol{\sigma}$, the scalar degradation factor $(1-d)$ is additionally employed
\begin{equation}\label{eq:CauchyStress}
    \boldsymbol{\sigma} = (1-d)\overline{\boldsymbol{\sigma}}.
\end{equation}
The computation of $d$ depends on $\overline{\boldsymbol{\sigma}}$ and a set of two hardening variables $\tilde{\boldsymbol{\varepsilon}}^{pl}$. These hardening variables are referring to equivalent plastic strains and are introduced to differentiate between behavior under compression (crushing failure; $\tilde{\varepsilon}_c^{pl}$) and tension (cracking failure; $\tilde{\varepsilon}_t^{pl}$). In addition to the degradation of the elastic stiffness, these also control the evolution of the yield function \citep{Jankowiak.2005}
\begin{equation}\label{eq:YieldFunctionSimple}
    F(\overline{\boldsymbol{\sigma}},\tilde{\boldsymbol{\varepsilon}}^{pl}) \leq 0 \text{ with: } \tilde{\boldsymbol{\varepsilon}}^{pl} = \left[ \substack{\tilde{\varepsilon}_c^{pl} \\ \tilde{\varepsilon}_t^{pl} } \right].
\end{equation}
$\overline{\boldsymbol{\sigma}}$ can be decomposed into one invariant describing the volume change (hydrostatic stress $\overline{p}$)
\begin{equation} \label{eq:hydrostatic}
    \overline{p} = -\frac{1}{3}\text{trace}(\overline{\boldsymbol{\sigma}})
\end{equation}
and one component describing the shape change (deviatoric stress $\overline{\textbf{S}}$) 
\begin{equation} \label{eq:deviatoric}
    \overline{\boldsymbol{S}} = \overline{\boldsymbol{\sigma}} + \overline{p}\textbf{I}
\end{equation}
with \textbf{I} as the identity tensor. $\overline{\boldsymbol{S}}$ is then transformed into the invariant von Mises equivalent stress $\overline{q}$
\begin{equation} \label{eq:vonMises}
    \overline{q} = \sqrt{1.5 (\overline{\boldsymbol{S}}:\overline{{\boldsymbol{S}}})}.
\end{equation}

The yield function $F$ is expressed in terms of $\overline{p}$, $\overline{q}$, and the Macauley Bracket $\langle x \rangle = 0.5(|x|+x)$ as
\begin{equation}
    F(\overline{p},\overline{q},\tilde{\boldsymbol{\varepsilon}}^{pl})=\frac{1}{1-\alpha} \left( \overline{q} + 3 \alpha \overline{p} + \beta \left( \tilde{\boldsymbol{\varepsilon}}^{pl}  \right) \langle \overline{\sigma}_{\rm max} \rangle - \gamma \langle - \overline{\sigma}_{\rm max} \rangle \right) - \overline{\sigma}_c \left( \tilde{\varepsilon}_c^{pl} \right) \geq 0 .
\end{equation}
Its shape is defined by $\alpha$, $\beta$ and $\gamma$, which are calculated according to
\begin{equation}\label{eq:alpha}
    \alpha = \frac{(\sigma_{b0}/\sigma_{c0})-1}{2(\sigma_{b0}/\sigma_{c0})-1{}},
\end{equation}
\begin{equation}\label{eq:beta}
    \beta = \frac{\overline{\sigma}_c(\tilde{\varepsilon}_c^{pl})}{\overline{\sigma}_t(\tilde{\varepsilon}_t^{pl})}(1-\alpha)-(1+\alpha),
\end{equation}
\begin{equation}\label{eq:gamma}
    \gamma = \frac{3(1-K_c)}{2K_c-1}.
\end{equation}
The ratio of initial equibiaxial to uniaxial compressive yield stress $\frac{\sigma_{b0}}{\sigma_{c0}}$ is used to compute $\alpha$ (eq. \eqref{eq:alpha}). The function $\beta$ considers the effective stress $\overline{\boldsymbol{\sigma}}$ computed by means of the hardening variables $\tilde{\varepsilon}^{pl}_c$ and $\tilde{\varepsilon}^{pl}_t$ (eq. \eqref{eq:beta}), while $\gamma$ is derived from $K_c$, which represents the ratio of the second stress invariant of the tensile meridian to that of the compressive meridian (eq. \eqref{eq:gamma}). $\frac{\sigma_{b0}}{\sigma_{c0}} = 1.16$ and $K_c=\frac{2}{3}$ are commonly used \citep{vanMeirvenne.2018}. $\overline{\sigma}_{\rm max}$ denotes the maximum eigenvalue of $\overline{\boldsymbol{\sigma}}$ \citep{Jankowiak.2005}.

For non-associative plasticity, the plastic flow $\dot{\boldsymbol{\varepsilon}}^{pl}$ results from the derivative of the flow potential function
\begin{equation}\label{eq:FlowPotential}
    \dot{\boldsymbol{\varepsilon}}^{pl} = \dot{\lambda} \frac{\partial G(\overline{\boldsymbol{\sigma}})}{\partial\overline{\boldsymbol{\sigma}}} \text{ with: } G(\overline{p},\overline{q}) = \sqrt{\epsilon f_{ctm} \tan(\psi)^2 + \overline{q}} - \overline{p} \tan(\psi).
\end{equation}
The Drucker-Prager hyperbolic function is used for this purpose. It also takes $\overline{p}$ and $\overline{q}$ as input parameters and is defined by the eccentricity $\epsilon$, the mean tensile strength of the concrete $f_{ctm}$ and the dilation angle in the $\overline{p}-\overline{q}$-plane $\psi$. Commonly used values are $\epsilon=0.1$\,mm and $\psi=30$\,° \citep{vanMeirvenne.2018}. 

The CDP distinguishes between compressive and tensile loading. For both loading paths, the formulation developed by \citep{Kratzig.2004} was used, employing the parameters suggested by \citep{Birtel.2006}. Details regarding the formulations can be derived from their work.  Under uniaxial compressive stress, ductile behavior with pronounced non-linear hardening and a softer drop in resistance is captured. As shown in Figure \ref{fig:ConstLaw}a, the first branch of the stress-strain relationship is linear-elastic $\sigma_{c(1)}(\varepsilon_c)$ (for $\sigma_c<0.4f_{cm}$). The second branch $\sigma_{c(2)}(\varepsilon_c)$ of the stress–strain relation represents the non-linear hardening phase (for $0.4f_{cm} \leq \sigma_c \leq f_{cm}$). It reflects progressive microcracking and damage accumulation in the cement matrix, accompanied by stress redistribution and aggregate interlock, which together enable a continued increase in load-carrying capacity despite a gradual reduction in stiffness \citep{vanMier.1987}. The third branch describes the behavior after the stress exceeds the mean compressive strength $f_{cm}$. It is governed by a progressive loss of load bearing capacity due to crack formation.

Under tension (Figure \ref{fig:ConstLaw}b), the model depicts brittle behaviour, which occurs after the mean tensile strength $f_{ctm}$ is exceeded. Subsequently, for $\sigma_t > f_{ctm}$, $\sigma_t$ is calculated from the crack width $w$, which according to \citep{Hillerborg.1983} is computed as $w = l_e \cdot \varepsilon^{in}_t$ (with the element length $l_e$).

The constitutive law of the steel is linear-elastic, since the prestress applied to the tendon $\sigma_{p0}$ in the experiments did not exceed the yield strength of St 1375/1570. The constitutive law can hence be described using only the Young's Modulus $E_p$ and Poisson Ratio $\nu_p$.

\subsubsection{Contact Formulation}\label{sec2.2.2}
\begin{figure}[h]
\centering
\includegraphics[width=0.9\textwidth]{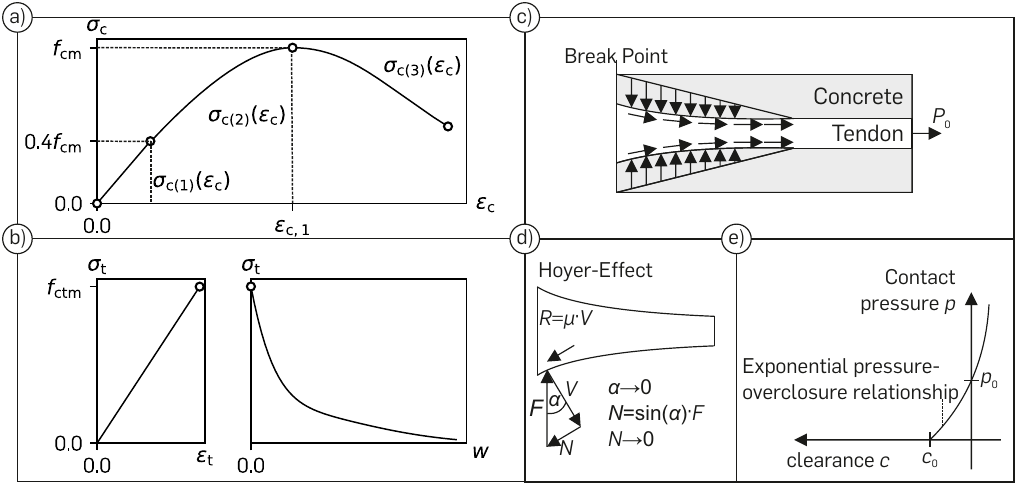}
\caption{Specifications of the computational model: behaviour of concrete under uniaxial compression (a), behaviour of concrete under uniaxial tension (b), close-up of the interface at the breakage point (c), local forces at the steel-concrete interface (d), contact specification normal to the interfaces by an exponential relationship between clearance $c$ and contact pressure $p$, according to \textit{ABAQUS} \citep{DassaultSystemesSimuliaCorporation.} (e).}
\label{fig:ConstLaw}
\end{figure}

The pretensioning of a tendon leads to a reduction of its diameter $\varnothing_p$. The release of the prestressing force during a tendon breakage hence results in an increase of $\varnothing_p$ to its initial value (see Figure \ref{fig:ConstLaw}c). To appropriately model this so called Hoyer effect \citep{Briere.2013}, the tendon is modeled as a volume element. Contact between steel and concrete is prescribed parallel and orthogonal to the concrete-tendon interface (Figure \ref{fig:ConstLaw}d). Normal to the interface, an exponential relationship provided in \textit{ABAQUS} is used. The contact pressure $p$ is computed from the clearance between the interfaces $c$ by an exponential function. It is defined by the clearance at which contact pressure is initiated ($c_0$) and the contact pressure $p_0$ when $c=0$, as is visualized in Figure \ref{fig:ConstLaw}e. The employed contact formulation allows a realistic load transfer, while concurrently permitting slip and decoupling as a consequence of local damage. 

Parallel to the interface, contact has been modeled by employing Coulomb's friction law through the friction coefficient $\mu$.

\subsubsection{Simulation Steps and Boundary Conditions}\label{sec2.1.3}
The simulation of the tendon breakage consists of three subsequent steps (cf. Table 1). The boundary conditions are set accordingly. Symmetry w.r.t. $yz$-plane at the breakage point ($x=0$) is used and only one part of the beam (free end at $l=1200$\,mm, cf. Figure \ref{fig2}) is modeled. In the symmetry plane ($x=0$), a corresponding boundary condition ($u_x=\theta_y=\theta_z=0$) is applied to the beam (concrete and tendons). It should be noted that solid elements do not possess rotational degrees of freedom, which reduces the symmetry conditions to $u_x=0$. Vertical displacements are prohibited ($u_z=0$) through all simulation steps (1-3) at the bottom of the beam ($z=100$\,mm), while displacements in $y$ ($u_y$) were not constrained.

To implement the prestress to the tendons, a predefined stress field ($\sigma_{p,0} =755 $\,N/mm²) in $x$-direction is initially applied to them. During the first simulation step (\textit{Initial}), axial displacements of the  tendons are prohibited at their ends ($x=0$ and $x=l$). Then, the initial boundary conditions are altered in the two subsequent simulation steps: In step 2 (\textit{Loading}), the transfer of the prestress $\sigma_{p,0}$ from the tendons to the concrete is simulated by releasing the fixation of $u_x$ at the free end ($x = l$) for all tendons. In simulation step 3 (\textit{Break}), the breakage is initiated by releasing the boundary condition of the central tendon at $x=0$. The computation of the strain difference between the simulations steps as shown in Figure \ref{fig:num_and_expResults} (right) corresponds to the measurements taken in the experiment. Table~\ref{tab:BoundaryConditions} summarizes the sequence of loading steps and the related boundary conditions.

\begin{table}[h]
	\centering
	\caption{Boundary conditions for different simulation steps of the computational model}
	\label{tab:BoundaryConditions}
	
	\begin{tabular*}{\textwidth}{@{\extracolsep{\fill}}lcccc}
		\toprule
		\shortstack{Simulation\\Step} &
		\shortstack{Bottom Surface\\($z=0.5h$)} &
		\shortstack{Central Tendon Cross\\Section ($x=0$)} &
		\shortstack{Concrete Cross\\Section ($x=0$)} &
		\shortstack{Tendons Free\\End ($x=l$)} \\
		\midrule
		Initial (1) & $u_z=0$ & $u_x=0$ & $u_x=0$ & $u_x=0$ \\
		Loading (2) & $u_z=0$ & $u_x=0$ & $u_x=0$ & [-] \\
		Breakage (3) & $u_z=0$ & [-] & $u_x=0$ & [-] \\
		\bottomrule
	\end{tabular*}
	
\end{table}

\subsection{Experimental Results}
\label{sec3.1}
\begin{figure}[h]
	\centering
	\begin{subfigure}[t]{0.45\textwidth}
		\centering
		\includegraphics[width=\linewidth]{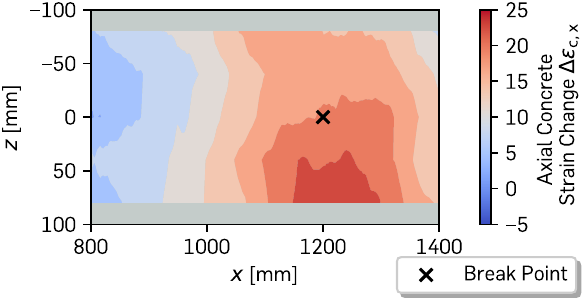}
	\end{subfigure}%
	\hfill
	\begin{subfigure}[t]{0.45\textwidth}
		\centering
		\includegraphics[width=\linewidth]{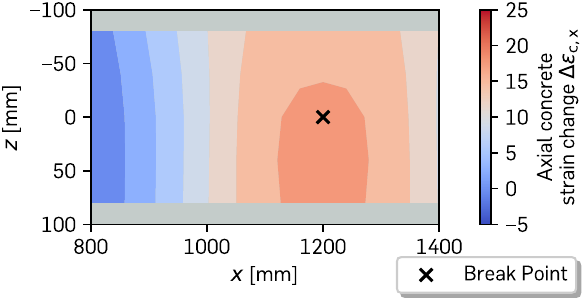}
	\end{subfigure}
	\caption{Strain field determined in the direction of the tendon axis $\Delta \varepsilon_{c,x}$. (Left) Experimentally, resulting from a broken tendon and (right) computational result as strain changes between simulation steps 2 and 3}
	\label{fig:num_and_expResults}
\end{figure}

The axial strains ($x$-direction) measured on the concrete surface with DFOS after the breakage of the tendon $\Delta \varepsilon_{c,x}$ are visualized in Figure \ref{fig:num_and_expResults} (left). The regions between the sensor strands were interpolated using bilinear interpolation \citep{Paul.2025}.  Positive strain changes of up to 25\,µm/m are measured in the proximity of the breakage point (marked with a black cross). As the distance from the breakage in $x$-direction increases, strains exhibit a decreasing trend, reaching a value of 0\,µm/m at approximately 350\,mm. Furthermore, a slight change in the strain field is observed in the $z$-direction. The strain changes exhibit a maximum at the bottom edge ($z = 0.5h$, $\Delta \varepsilon_{c,x}\approx 25$\,µm/m) 
while smaller changes are measured at the upper edge ($z = -0.5h$, $\Delta \varepsilon_{\rm c,x} \approx18$\,µm/m). This can be attributed to the full-surface support of the specimen on its bottom surface ($z = 100$\,mm) during the experiment and corresponding friction.

It is important to note that the DFOS were applied after the induction of the prestressing force (after step 2), resulting in the specimen being compressed at the reference time $t_0$. Due to the tendon breakage (at $t_1$), the compression diminishes around the breakage point and increases again towards the free end because of the re-anchorage of the tendon in the surrounding concrete~\citep{Hegger.2010}. The measured tensile strain changes are therefore calculated as ($\Delta \varepsilon_{c,x} = \varepsilon_{c,x}(t_1)-\varepsilon_{c,x}(t_0)$). 

\subsection{Numerical Results}
\label{sec3.2}
Figure \ref{fig:num_and_expResults} (right) shows the strain changes coming from the simulation model presented in section~\ref{sec2.2} at simulation step 3. The strain changes are computed at positions corresponding to those of the sensors on the test specimen (subsection~\ref{sec2.2.1}). To ensure better comparability with the measurements, the $x$-coordinates of the strain field were shifted to match those of the experimental investigation. Local tensile strains of up to 20\,µm/m are observed around the breakage point. They decrease with increasing distance in $x$-direction. As observed in the experiment, also in the simulation slightly greater strain changes are computed at the bottom of the beam ($z = 100$\,mm; 20\,µm/m) than at the top ($z = -100$\,mm; 17\,µm/m). This can be attributed to the implementation of the boundary condition on the bottom surface (see Table~\ref{tab:BoundaryConditions}). Overall, the calculated strain field is slightly smoother than the experimental one, as the influence of measurement noise and minor inaccuracies in sensor localization during the touch-to-locate process is mitigated.

\subsection{Inverse problem definition}
For the inverse problem, the parameters to be inferred are $\boldsymbol{\theta} = [E_{cm},\, p_{0},\, c_{0},\, \mu]$. All the other parameters detailed in Sections \ref{sec2.1} and \ref{sec2.2} required for the numerical model are set to their default values as identified from literature or material testing (see Table~\ref{tab:parameters}). The geometry is set to reproduce the experimental set-up described in Section \ref{sec2.1}.

\begin{table}[htbp]
\centering
\caption{Model parameters, descriptions, units, and default values. Parameters calibrated in $\bm{\theta}$ are indicated.}
\label{tab:parameters}
\begin{tabular}{llp{5.2cm}cc}
\hline
\textbf{Characteristic} & \textbf{Parameter} & \textbf{Description} & \textbf{Unit} & \textbf{Default} \\
\hline
\multirow{8}{*}{Concrete}
 & $E_{cm}$  & Young's modulus of concrete & MPa & Calibrated \\
 & $\nu_c$   & Poisson's ratio of concrete & -- & 0.20 \\
 & $\sigma_{b0} / \sigma_{c0}$ & Ratio of initial equibiaxial to uniaxial compressive yield stress & -- & 1.16 \\
 & $K_c$ & Ratio of second stress invariants of tensile and compressive meridians & -- & $2/3$ \\
 & $\epsilon$ & Eccentricity of CDP & mm & 0.1 \\
 & $f_{ctm}$ & Mean tensile strength of concrete & MPa & 4.1 \\
 & $\psi$    & Dilation angle of CDP & $^\circ$ & 30.0 \\
 & $\eta$ & Viscosity Parameter & -- & 0.0005 \\
\hline
\multirow{1}{*}{Steel}
& $E_p$ & Young's modulus of steel & MPa & 196000 \\
 & $\nu_p$   & Poisson's ratio of steel & -- & 0.30 \\
\hline
\multirow{2}{*}{Contact}
 & $\mu$     & Friction coefficient & -- & Calibrated \\
 & $c_0$ & Clearance at which contact pressure is initiated & mm & Calibrated \\
 & $p_0$ & Contact pressure at clearance $c=0$ & MPa & Calibrated \\
\hline
\multirow{1}{*}{Solver}
 & $\zeta$   & Numerical damping for solver stability & -- & 0.0002 \\
\hline
\multirow{1}{*}{Error}
 & $\sigma$  & Sensor noise standard deviation & -- & 0.01 \\
\hline
\end{tabular}
\end{table}

The observations used for the calibration are a discretization of the observations of the DFOS set-up described in Section \ref{sec2.1}. Five rows of measurements are observed at heights $z \in \lbrace-80, -40, 0, 40, 80\rbrace$~mm from the middle line of the beam with 11 measurements each starting at the tendon breakage point towards the end of the beam ($x=0$) with a distance of $40$~mm. In total, 55 measurement points are collected.

The predictions of the FEM model will be used as output to be calibrated against. These predictions provide the difference in tensile strain $\Delta\varepsilon_{c,x}$ before and after the breakage at the location of the DFOS. The predictions are collected at the equivalent locations in the model as the observations. The aim is to calibrate the model parameters such that the FEM model can reliable reproduce the observations.

\subsection{Gaussian Process Training}
\label{sec3.3}
The Gaussian Process (GP) surrogate model was trained using four input parameters: $E_{cm}$, $c_{0}$, $p_{0}$, and $\mu$. These parameters were selected following a sensitivity analysis in \citep{Paul.2026}, which demonstrated their dominant influence on the system response, while the remaining parameters exhibited negligible effects. The output quantity of interest was the difference in the strain field $\Delta\varepsilon_c$, extracted at $N_{\text{points}}$ spatial locations over the surface of the finite element (FEM) model corresponding to the DFOS setup. This resulted in an output vector of dimension $N_{\text{points}}$ for each simulation run.

A total of $N_{\text{samples}} = 100$ training points were generated using Latin Hypercube Sampling (LHS), ensuring good coverage of the multidimensional input space. The corresponding FEM simulations provided the training outputs. The GP was implemented with a Radial Basis Function (RBF) kernel combined with a white noise kernel, using the \texttt{scikit-learn} framework. Inputs were normalized to the range of 0 to 1 in the space of the training samples. Outputs were normalized jointly as well to [0,1] over the whole dataset of training points. Table~\ref{tab:gp_ranges} summarizes the ranges of the input parameters and the kernel hyperparameters.

\begin{table}[h]
	\centering
\caption{Input parameter ranges and Gaussian Process hyperparameter bounds.}
\label{tab:gp_ranges}
\begin{tabular}{lcc}
\hline
\textbf{Parameter} & \textbf{Lower Bound} & \textbf{Upper Bound} \\
\hline
$c_0$ [mm] & $0.2012$ & $0.7988$ \\
$p_0$ [MPa] & $2.008$ & $5.992$ \\
$E_{cm}$ [MPa] & $2.702e+04$ & $3.898e+04$ \\
$\mu$ [-] & $0.202$ & $1.198$ \\
\hline
\text{Length scales} $\ell_i$ & $0.01$ & $100.0$  \\
\text{Signal variance} $\sigma_f^2$ & $0.001$ & $1000.0$  \\
\text{Noise variance} $\sigma_n^2$ & $1e-07$ & $0.1$  \\
\hline
\end{tabular}
\end{table}

The hyperparameters were optimized by maximizing the log-marginal likelihood. The noise variance $\sigma_n^2$ converged to the minimum admissible value ($\sigma_n^2 = \sigma_{n,\min}^2$), which reflects the high consistency between the surrogate and the training data. One GP is trained for each of the observation points, leading to different optimal hyperparameters for each case.

The predictive performance of the trained Gaussian Process (GP) model was assessed using an independent validation set of $N_{\text{val}}$ samples generated through Latin Hypercube Sampling within the same parameter ranges as the training data. The FEM simulations corresponding to these validation points served as the reference values for comparison. The evaluation was based on standard regression metrics, including the coefficient of determination ($R^2$), the root mean square error (RMSE), the mean absolute error (MAE), the maximum error, the normalized RMSE (NRMSE) in percentage, and coverage statistics based on the absolute $Z$-value. The results were validated on 49 random samples on the training domain and they are summarized in Table~\ref{tab:gp_validation_metrics}.

\begin{table}[h]
	\centering
\caption{Training and validation metrics for both GP emulators.}
\label{tab:gp_validation_metrics}
\begin{tabular}{lcc}
\hline
\textbf{Metric} & \textbf{Train}  & \textbf{Validation} \\
\hline
RMSE & 0.1627  & 0.4409 \\
MAE & 0.1127 & 0.2738 \\
R2 & 0.9991 & 0.9940 \\
Max Error & 0.8214 & 1.9970 \\
NRMSE (\%) & 0.67 & 1.74 \\
Mean |z| & 0.6277 &  0.7481 \\
|z| < 2 (\%) & 94.98 &  93.21 \\
|z| > 0.5 (\%) & 43.48 & 52.39 \\
\hline
\end{tabular}
\end{table}

The obtained $R^2$ value close to unity confirmed that the GP surrogate accurately reproduced the FEM reference outputs across the parameter space. The RMSE and MAE values were small relative to the magnitude of the strains $\varepsilon$, indicating low overall prediction error. Additionally, the predictive uncertainty estimates provided by the GP were consistent with the residuals observed in the validation set. The distribution of the normalized prediction errors remained within the $95\%$ confidence bounds for the majority of the validation points, demonstrating that the model not only achieved high accuracy but also provided reliable uncertainty quantification. These results validate the suitability of the GP surrogate for representing the input–output mapping of the FEM model in the DFOS configuration.

\subsection{Parameter Updating}
\label{sec3.4}
The Bayesian calibration is carried out using the affine-invariant ensemble sampler implemented in \texttt{emcee}, with $N_{\text{walkers}}=20$ walkers evolved for $N_{\text{iter}}=10000$ iterations. The parameter space is explicitly bounded to remain within the domain of validity of the Gaussian Process (GP) surrogate. In the presence of residual model–data discrepancy and strong parameter bounds, the resulting posterior may exhibit multiple disconnected modes, some of which correspond to low-probability regions near the surrogate limits. In such cases, ensemble samplers may converge to distinct local probability wells \citep{Hou2012}.

To address this issue, we apply a likelihood-based clustering and pruning procedure after burn-in and at the end of sampling, designed to retain the dominant posterior mode while discarding walkers trapped in low-probability regions. The full algorithmic formulation is provided in Appendix~\ref{ap:clustering-pruning}; here we summarize the key principles. After burn-in, the mean log-posterior of each walker $i$ is computed over the final fraction $\alpha$ of its trajectory
\begin{equation}
\ell_i = \frac{1}{T}\sum_{t=1}^{T} \ln \pi_{i,t},
\qquad T=\lfloor \alpha,T_{\mathrm{burn}}\rfloor .
\end{equation}
The values $\lbrace\ell_i\rbrace_{i=1}^{N_\text{walkers}}$ are sorted in ascending order, and clusters are identified by detecting large jumps in consecutive differences
\begin{equation}
d_k = \ell_{(k+1)} - \ell_{(k)}, \qquad k=1,\dots,N_{\mathrm{walkers}}-1,
\end{equation}
where a jump is declared whenever
\begin{equation}
d_k > \gamma,\mathrm{median}{d_j},
\end{equation}
with $\gamma=5$. These jumps partition the ensemble into disjoint clusters in log-posterior space. The cluster corresponding to the largest coherent posterior mass is retained, while walkers belonging to lower-probability clusters are pruned. Pruned walkers are replaced by new initial states obtained via convex combinations of retained walkers, ensuring that the ensemble remains within the high-probability region and inside the GP training domain. The corrected ensemble is then evolved for an additional $N_\text{iter}=10000$ iterations. At the end of the full chain, the same clustering criterion is reapplied to obtain the final posterior ensemble, without resampling. The likelihood function is Gaussian and evaluated using the GP surrogate mean predictions and the observed DFOS data supposing a Gaussian noise variance $\sigma=0.01$. GP hyperparameters are fixed during MCMC sampling at the optimized values obtained in Section~\ref{sec3.3}.

Two updating approaches were considered. In the first approach, the parameter vector was defined as $\boldsymbol{\theta} = [E_{cm},\, p_{0},\, c_{0},\, \mu]$. In the second approach, the parameter vector was augmented with $\sigma_{E_{cm}}$, where $E_{cm}$ is interpreted as the mean of a lognormal distribution and $\sigma_{E_{cm}}$ as its standard deviation, such that $\tilde{E}_{cm} \sim \mathcal{LN}(E_{cm}, \sigma_{E_{cm}})$. The extended parameter $\tilde{E}_{cm}$ was propagated through the GP surrogate using a PCE of degree~2, as described in Section~\ref{sec2.4}. The likelihood function was adapted accordingly, as shown in Equation~\ref{eq:embedded_likelihood}.

\begin{figure}[h]
	\centering
	\begin{subfigure}[t]{0.36\textwidth}
		\centering
		\includegraphics[width=\textwidth]{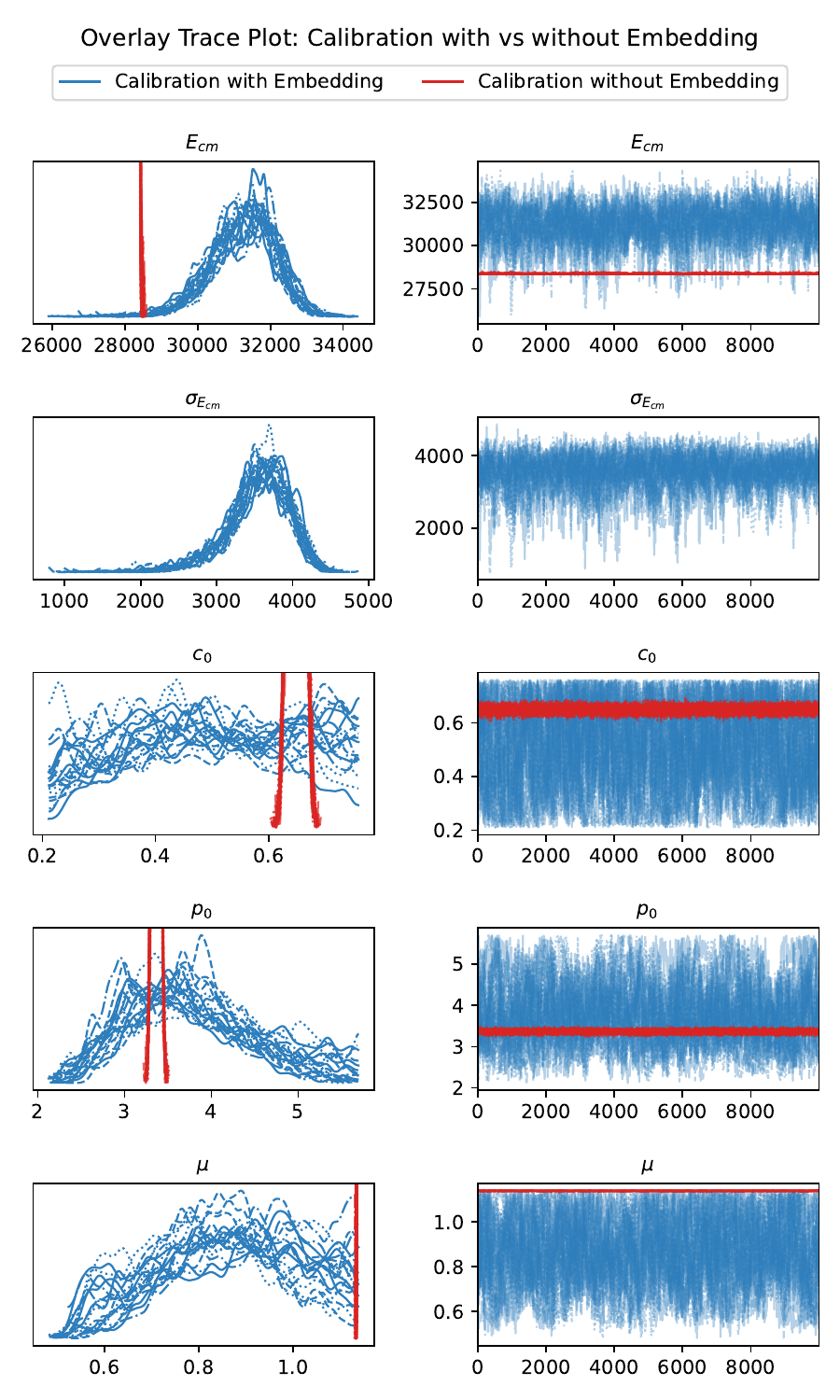}
		\caption{Trace plot of MCMC calibration. Without (in red) and with embedding (in blue).}
		\label{fig:trace_plot_mcmc_non_uq}
	\end{subfigure}%
	\hfill
	\begin{subfigure}[t]{0.59\textwidth}
		\centering
		\includegraphics[width=\textwidth]{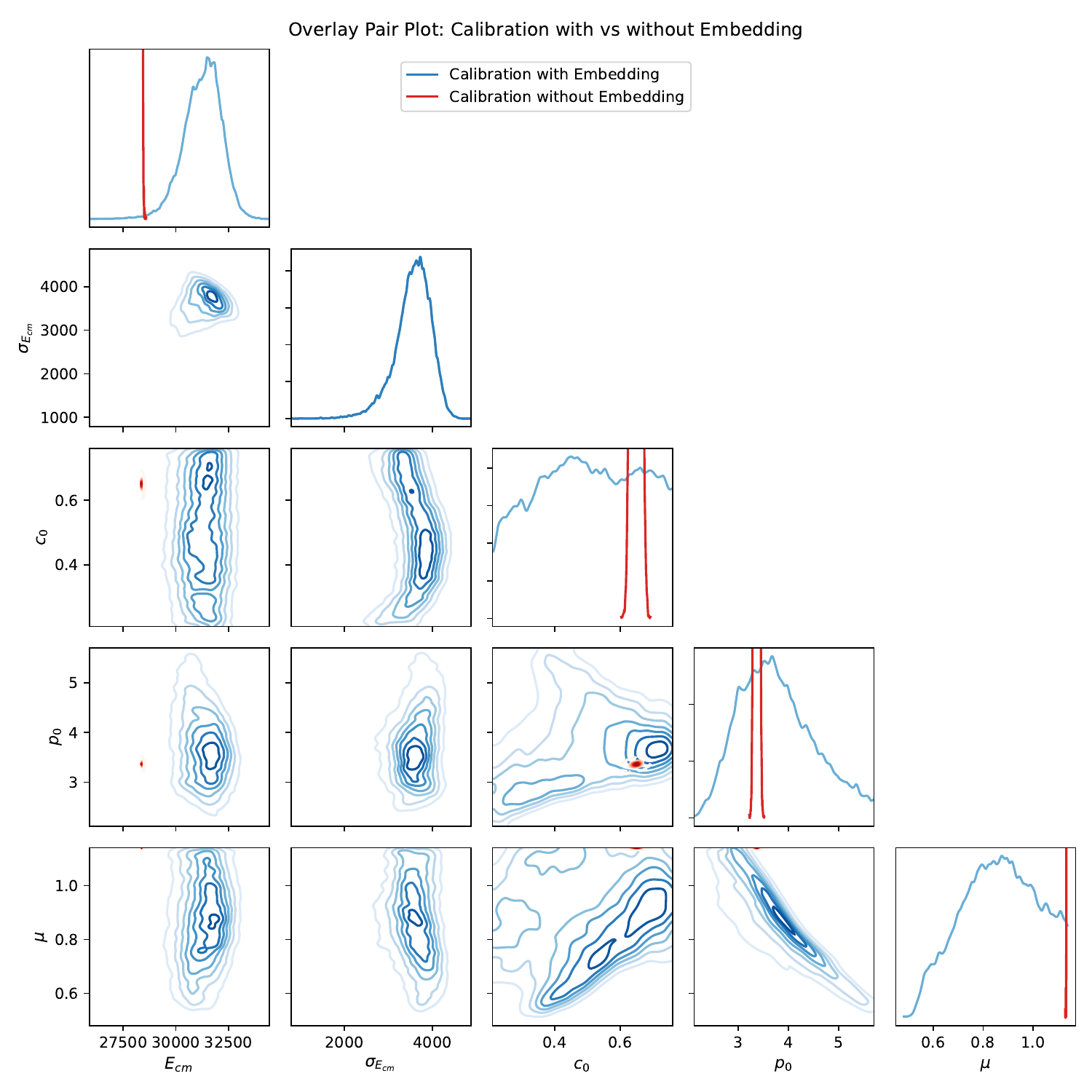}
		\caption{Pair plot of MCMC calibration. Without (in red) and with embedding (in blue). PDF comparison in the diagonal.}
		\label{fig:pair_plot_mcmc_non_uq}
	\end{subfigure}
	\caption{Comparison with and without embedding of MCMC calibration for the parameters $E_\mathrm{cm}$, $\sigma_{E_{cm}}$, $p_0$, $c_0$, and $\mu$. The left subfigure shows trace plots, and the right shows parameter pair plots}
	\label{fig:mcmc_results_combined_non_uq}
\end{figure}

\begin{figure}[h]
\centering
\includegraphics[width=0.9\textwidth]{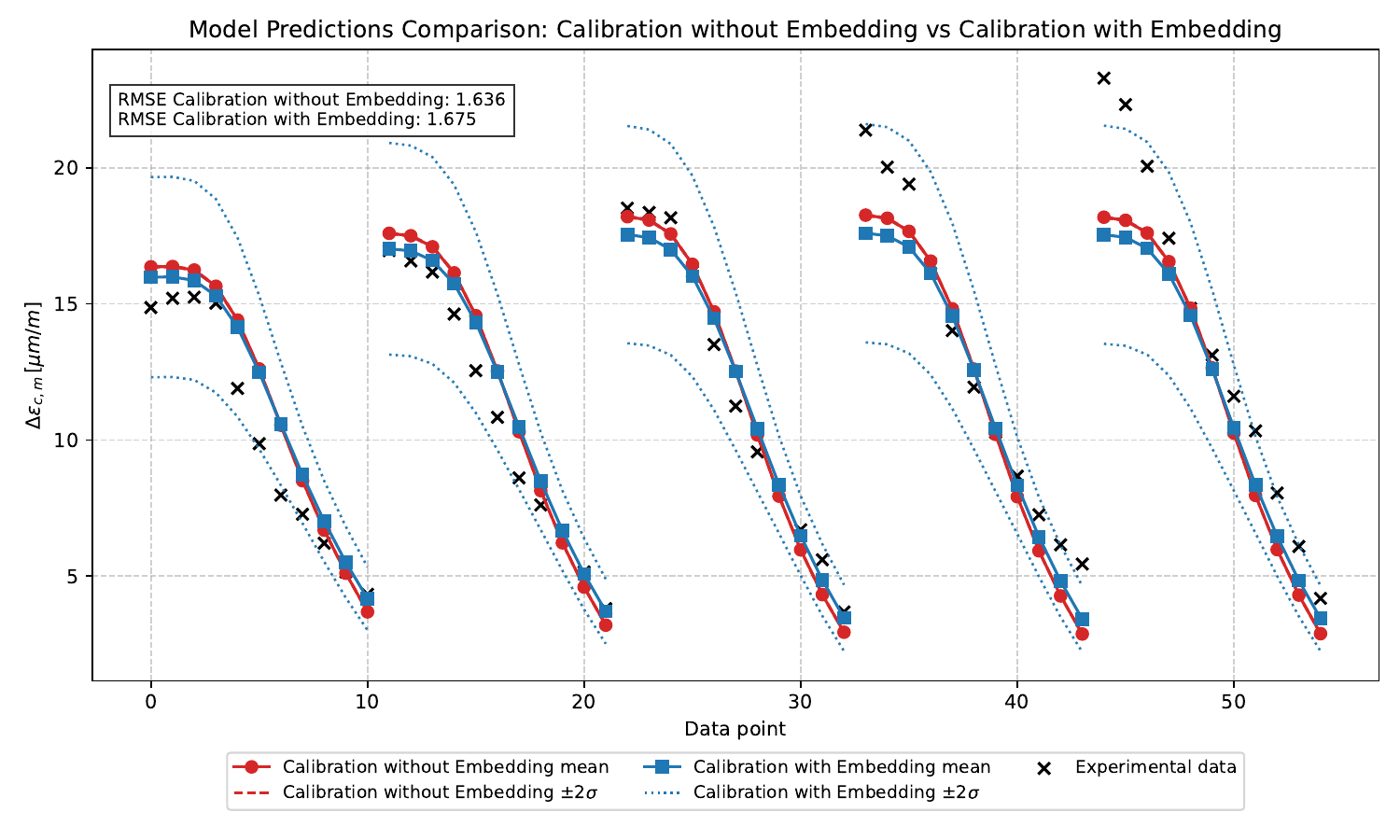}
\caption{Predicted confidence intervals at $95\%$ of the posterior predictive for $\hat{\tilde{\bm{\theta}}}$ with and without embedding. The confidence interval in the calibrated model without embedding depends only on the prescribed noise $\sigma$.}
\label{fig:prediction_vs_data_non_uq}
\end{figure}

The comparison between the non-embedded and embedded uncertainty calibration results is summarized in Table~\ref{tab:predictive_statistics}, while the calibration performance metrics are reported in Table~\ref{tab:triple_comparison}. The corresponding trace and pair plots of the posterior samples are shown in Figure~\ref{fig:mcmc_results_combined_non_uq}, and the predictive comparisons against the experimental observations are presented in Figure~\ref{fig:prediction_vs_data_non_uq}. Without embedding the uncertainty, the confidence intervals (CIs) are generated exclusively from the prescribed noise $\sigma$. Despite this, the resulting CIs fail to encompass the experimental observations, indicating the presence of model-form uncertainties that are not captured by the GP variance alone. As a consequence, the calibrated parameters are driven toward the boundaries of their prior ranges, reflecting an artificial compensation for the missing model discrepancy. This behaviour indicates that the calibration in this configuration is not physically meaningful, as reflected by the value of $\mu$ tending toward its maximum allowable value and $E_{cm}$ approaching its minimum. In this case, the predictive variance does not constitute an actual quantification of MFU in the simulation model, but rather a by-product of uncertainty in the GP surrogate, which becomes less accurate near the boundaries of the domain due to the sampling scheme used to generate the training dataset.

When the uncertainty is embedded through the stochastic representation of $E_{cm}$, the inferred variance is directly associated with the material stiffness. This embedding inflates the CIs and improves their coverage of the observational data. The posterior distributions of the parameters are observed to be less concentrated, as a result of the flatter likelihood induced by the inclusion of variance in the quantification. Nevertheless, these posteriors converge toward more reasonable values than the extreme estimates obtained in the model without embedding. In particular, the inferred Young’s modulus of approximately 31000~MPa, with an associated uncertainty of 3500~MPa, lies within the expected range of experimentally observed values. Additionally, a general lack of influence of $d$, together with a clear correlation between $p_0$ and $\mu$, can be observed in the posterior behaviour. The $Z$-values, defined as the normalized residuals between observations and model predictions, confirm these findings. They show a clear increase in coverage (87\% with versus 60\% without embedding) and a decrease in outliers (13\% of absolute $Z$-values exceeding 2 with versus 40\% without embedding), despite an increase in the percentage of the dataset for which the discrepancy is smaller than the predictive intervals (42\% with versus 16\% non-without embedding). Nevertheless, some observations still fall outside the 95\% predictive intervals, indicating that part of the model-form uncertainty remains unaccounted for, even after embedding. Moreover, the predicted variance is significantly larger than the observed residual in more than one third of the predictions, suggesting a general overestimation of the predictive uncertainty due to the remaining discrepancies.

\begin{table}[h]
\centering
\caption{Priors for parameters used in MCMC calibration.}
\label{tab:priors}
\begin{tabular}{l l l}
\hline
Parameter & Prior type & Parameters / bounds \\
\hline
$E_{cm}$ & Lognormal & $\mu=33000\text{ MPa}, \sigma=3300\text{MPa}$ $[25200,37050]$ \\
$\sigma_{E_{cm}}$ & Uniform & $[0.25,7.41]$\\
$p_{0}$ & Uniform & $[2.1,5.7]$ \\
$c_{0}$ & Uniform & $[0.21,0.76]$ \\
$\mu$ & Uniform & $[0.21,1.14]$ \\
\hline
\end{tabular}

\end{table}
\begin{table}[h]
\centering
\caption{Predictive statistics (residual-based and $|Z|$-values) for the non-embedded and embedded uncertainty calibration cases. Best values in bold}
\label{tab:predictive_statistics}
\resizebox{\textwidth}{!}{%
\begin{tabular}{l c c c c c c c c c c c}
\toprule
& \multicolumn{4}{c}{\textbf{Residual-based}} & \multicolumn{7}{c}{\textbf{$|Z|$-values}} \\
\cmidrule(lr){2-5}\cmidrule(lr){6-12}
\textbf{Case} & Mean & RMSE & Median & MAD & Mean & Std. Dev. & Median & MAD & \shortstack{$|Z|$ \\ $>2$ (\%)} & \shortstack{$|Z|$ \\ $<0.5$ (\%)} & \shortstack{Coverage \\ (95\%)} \\
\midrule
Non-UQ & 0.23 & \textbf{1.64} & 0.16 & 1.66 & 1.87 & 1.49 & 1.52 & 1.22 & 40 & \textbf{16} & 60 \\
UQ & \textbf{0.27} & 1.67 & \textbf{0.11} & \textbf{1.35} & \textbf{0.94} & \textbf{0.79} & \textbf{0.61} & \textbf{0.68} & \textbf{13} & 42 & \textbf{87} \\
\bottomrule
\end{tabular}
}
\end{table}
\begin{table}[h]
\centering
\caption{Comparison of posterior summaries acros calibrations with and without embedding.}
\label{tab:triple_comparison}
\resizebox{\textwidth}{!}{%
\begin{tabular}{l c c c c c }
\hline
& \multicolumn{2}{c}{Without embedding} & \multicolumn{2}{c}{With embedding}  \\
\cline{2-5}
Parameter & Mean (std) & 95\% CI & Mean (std) & 95\% CI \\
\hline
$E_{cm}$ & 28368.30 (18.35) & [28350.47, 28417.44] & 31244.27 (894.13) & [29336.98, 32803.68] \\
$\sigma_{E_{cm}}$ & -- & -- & 3548.81 (416.91) & [2581.66, 4216.89]\\
$p_{0}$ & 3.36 (0.03) & [3.30, 3.43] & 3.77 (0.73) & [2.56, 5.37]\\
$c_{0}$ & 0.65 (0.01) & [0.63, 0.67] & 0.50 (0.15) & [0.23, 0.75] \\
$\mu$ & 1.14 (0.00) & [1.14, 1.14] & 0.87 (0.15) & [0.58, 1.12] \\
\hline
\end{tabular}
}
\end{table}

\clearpage

To further investigate the remaining discrepancies observed in the predictions, an influence analysis was performed on the model calibrated with embedded uncertainty. The predictions, which represent the change in normal strains after breaking the tendon on the DFOS, still do not fully cover the experimental observations. Since the sensor data originate from five different DFOS located at distinct vertical positions $z$, the observations were grouped by their distance to the tendon break, forming subsets $S_i$. Each subset $S_i$ thus represents measurements obtained at a similar structural region with respect to the failure point.

\begin{figure}[h]
\centering
\includegraphics[width=\textwidth]{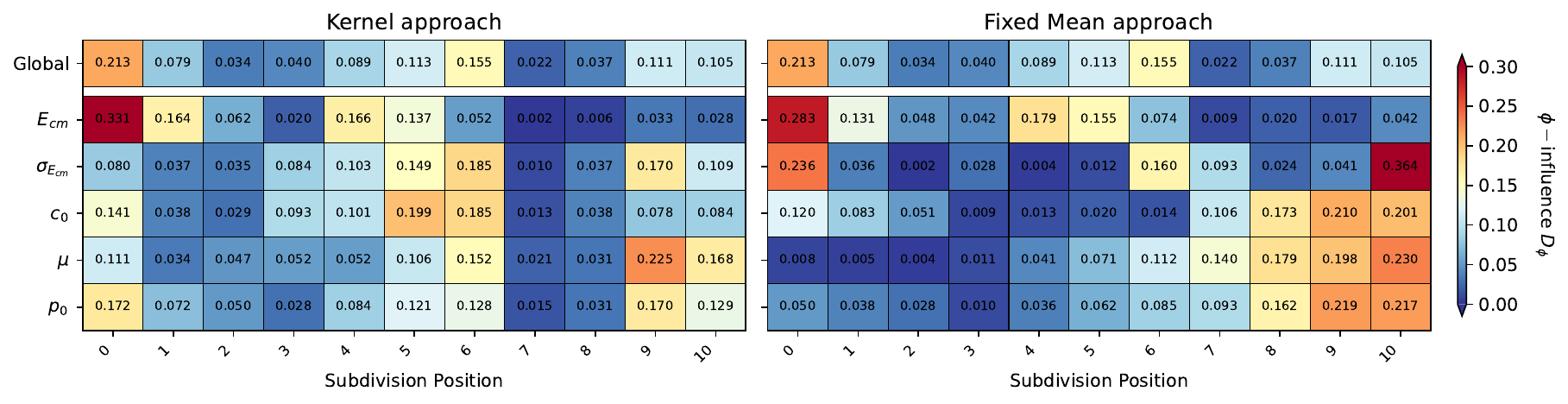}
\caption{Influence analysis on the model calibrated with embedded parameters. At the top for both sides, $\hat{D}_\phi(S_i)$ for the full posterior for each $S_i$. On the left, influence $\hat{D}^{(j)}_\phi(S_i)$ of each $S_i$ on the marginal posterior of the parameters $\theta^{(j)}$ using  KDE smoothing. On the right, influence $\hat{D}^{(j),(\text{fix})}_\phi(S_i)$ of each $S_i$ on the marginal posterior of the parameters $\theta^{(j)}$ fixing the means of the other parameters. Influence values have been normalized to add to 1 in their respective case to enable the direct comparison.}
\label{fig:influence_analysis}
\end{figure}

The influence of each subset $S_i$ on the posterior distribution of the parameters was quantified using $\phi$-divergences, specifically the reverse Kullback–Leibler (KL) divergence between the posterior obtained using all observations for calibration and only a subset of them (see Section \ref{sec2.5}). This approach allows assessing how the removal of each subset affects the posterior, thereby identifying which experimental groups exert the largest influence on the calibrated model. The resulting influence metrics are illustrated in Figure~\ref{fig:influence_analysis}. The top plots display the global influence $\hat{D}_\phi(S_i)$ for each subset on the full posterior, while the lower panels show, respectively, the marginal influences $\hat{D}^{(j)}_\phi(S_i)$ and $\hat{D}^{(j),(\text{fix})}_\phi(S_i)$ for each parameter $\theta^{(j)}$, obtained using two alternative regularization approaches. All influence values have been normalized such that they sum to one within each case, enabling direct comparison between subsets and parameter-wise influences.

The global influence analysis reveals a dominant contribution from the first subset, corresponding to the portion of the DFOS located closest to the tendon break. This region coincides with the observations that are least well captured by the predictive intervals, confirming that the local response near the failure point drives the largest posterior changes. A secondary effect is also observed in the sixth to seventh subdivision, affected by the increased discrepancies in the first set of observations (data points 7 and, to a lesser extent, 6 ).

The marginal influence analyses highlight that the parameter $E_{cm}$ is primarily driven by observations located close to the tendon breaking point. These measurements correspond to the region exhibiting the largest model–data discrepancies and the highest sensitivity to calibration, indicating that variations in $E_{cm}$ induce the most significant changes in the predicted strain response in this zone. The embedded variance parameter $\sigma_{E_{cm}}$ is also strongly affected in the fixed-point influence analysis, suggesting that part of the local discrepancy is being absorbed through an increased variability of the embedded stiffness rather than through systematic shifts in the mean response. A similar, albeit weaker, effect is observed for the last observation, which also displays residual discrepancy at the identified optimum.

Two complementary influence formulations are considered to disentangle these effects. In the kernel-based influence analysis, the parameters are marginalized, such that the influence measure reflects the sensitivity of the joint posterior distribution after integrating out parameter dependencies. In this formulation, part of the discrepancy attributed to a given parameter can be compensated by correlated variations in other parameters, leading to a more spatially homogeneous influence pattern for $\sigma_{E_{cm}}$. In contrast, the fixed-mean influence analysis evaluates sensitivity by conditioning on fixed values of all parameters except the one under investigation. This isolates the direct contribution of individual parameters to the posterior change and is therefore more sensitive to localized discrepancies that cannot be alleviated through compensating parameter interactions.

The KDE-based analysis provides a global view of how observations influence the inferred uncertainty when all parameter interactions are accounted for, while the fixed-mean analysis reveals which parameters are directly responsible for reconciling local discrepancies in the predictions. The latter is particularly relevant in the context of parameter transfer, where mean values are typically propagated to simulations of real structures. Taken together, the two influence measures indicate that the observed discrepancies near the breakage region stem from a localized model deficiency, which is partially absorbed through increased embedded stiffness variability when parameter interactions are allowed.

\subsection{Identifiability assessment of parameters of interest}
\label{sec3.5}

The final step of the analysis aims to assess the identifiability of parameters of interest when the calibrated and uncertainty-embedded model is transferred to a full-scale, realistic structure. The parameters inferred from the calibration with uncertainty quantification, specifically the stochastic Young's modulus $\tilde{E}_{cm}$, are now propagated to a new FE model representing a structural component with a different cross-section; it is based on the dimensions of a real bridge. This transfer enables the propagation of both epistemic and aleatory uncertainties obtained from the experimental calibration into a context that more closely resembles a real engineering application.

The motivation for this analysis arises from the intended use of the model in structural health monitoring of prestressed concrete bridges. In a realistic setting, the calibrated parameters from laboratory-scale experiments would be employed to interpret field measurements, with the objective of diagnosing potential failures in the prestressing tendons. In particular, if a tendon breaks, the resulting strain field on the concrete surface will change in a way that depends on the depth of the breakage $a$ \citep{Paul2024}. Identifying which tendon has failed therefore corresponds to solving an inverse problem: given a new set of observed strains, determine the value of $a$ (i.e., the depth of the broken tendon) that most likely produced them. Such identification is only feasible if the predictive distributions corresponding to different tendon depths are sufficiently distinct. The separability analysis presented here thus evaluates whether the embedded model uncertainty allows discrimination between potential tendon breakage depths.

To perform this assessment, the stochastic Young's modulus $\tilde{E}_{cm}$, characterized by its mean and variance from the calibration, is treated as a random variable and propagated through the FE model for each tendon depth $a$. This yields predictive distributions of the strain response $\varepsilon \sim \mathcal{G}_{\mathbf{x}}(a)$ at spatial points $\mathbf{x}$, quantifying how uncertainty in material and bond parameters translates into uncertainty in the observable strain field. 

\begin{figure}[h!]
	\centering
	\includegraphics[width=150mm]{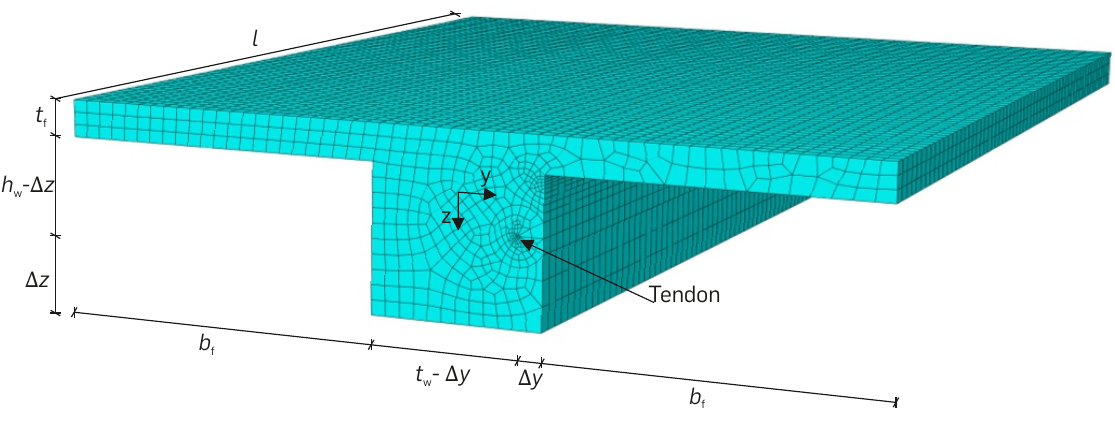}
	\caption{Computational model of an upscaled T-beam used to investigate the separability of predictions for different tendon depths}
	\label{fig:UpscaledModel}
\end{figure}

The upscaled structural model represents a symmetric T-shaped beam of length $l$, shown in Figure~\ref{fig:UpscaledModel}. Its cross-section is defined by the web height $h_w=1000$\,mm, web width $t_w=1200$\,mm, flange thickness $t_f=250$\,mm, and flange width $b_f=2250$\,mm. The geometric parameters remain constant for all analyses. Since the study focuses on the breakage of a single tendon and the concrete remains uncracked, the intact tendons are not explicitly modeled. The position of the tendon within the cross-section is defined by $(\Delta y, \Delta z)$, where $\Delta y$ denotes the embedding depth w.r.t to the evaluated concrete surface and corresponds to the control parameter $a$. The strain response is evaluated along the outer surface of the web at $y = 0.5 t_w$.

The central question of this study is whether the propagated uncertainty from the calibrated parameters still permits distinguishable predictions for different values of $a$. If the predictive distributions for different tendon depths overlap significantly, then new observations would not allow reliable inference of which tendon has failed. Conversely, well-separated predictive distributions imply that the distributed surficial strain measurement could be effectively used to identify the failure location. Hence, the separability of the predictive strain distributions provides a quantitative measure of the model’s diagnostic capability and defines a metric for the informativeness of potential sensor locations.

Direct evaluation of the FE model for all $a$ values and sensor points would be computationally prohibitive. To overcome this limitation, the propagation of $\tilde{E}_{cm}$ was performed using a Polynomial Chaos Expansion (PCE) of degree 2, yielding the mean $\tilde{\mathbf{m}}_{\mathrm{PCE}}$ and standard deviation $\tilde{\bm{\sigma}}_{\mathrm{PCE}}$ fields at each spatial point. Two Gaussian Process (GP) surrogate models were trained to emulate the PCE outputs: one for the mean field and one for the standard deviation. Each GP takes as input $(\mathbf{x}, a)$ and outputs either $\tilde{\mathbf{m}}_{\mathrm{PCE}}$ or $\tilde{\bm{\sigma}}_{\mathrm{PCE}}$. Both employ a isotropic Matern kernel with a white-noise component with inputs and outputs scaled to [0,1]. The optimized hyperparameters of the GP surrogates are listed in Table~\ref{tab:gp_hyperparameters_two}. The input dataset is composed of $x$, $z$ and $a$ in a grid $[5\times15\times10]$ divisions, from which 2500 data points are taken randomly for training the GPs and the remaining 5000 are used for validation. Validation metrics are presented in Table~\ref{tab:gp_validation_combined}.

\begin{table}[h]
	\centering
\caption{Hyperparameter summary for mean and std GP surrogates (input parameter $a$).}
\label{tab:gp_hyperparameters_two}
\begin{tabular}{lccc}
\hline
 & Bounds & \textbf{Mean GP} & \textbf{Std GP} \\
\hline
$\ell$ & $[0.01, 100.0]$ & 0.5561 & 0.2561 \\
$\sigma_f^2$ & $[0.001, 10000.0]$ & 780.3 & 0.6031 \\
$\sigma_n^2$ & $[1e-08, 0.01]$ & 1.373e-05 & 7.549e-05 \\
\hline
\end{tabular}
\end{table}
\begin{table}[h]
\centering
\caption{Validation metrics for GP mean and std emulators.}
\label{tab:gp_validation_combined}
\begin{tabular}{lcc}
\hline
\textbf{Metric} & \textbf{Mean GP} & \textbf{Std GP}\\
\hline
RMSE & 0.1639 & 0.0282\\
MAE & 0.05257 & 0.00938\\
R2 & 0.9978 & 0.9938\\
Max Error & 3.772 & 0.4837\\
NRMSE (\%) & 0.23 & 0.45\\
Mean |z| & 0.3125 & 0.2445\\
|z| < 2 (\%) & 97.70 & 97.51\\
|z| > 0.5 (\%) & 13.09 & 10.19\\
\hline
\end{tabular}
\end{table}

The separability of the predictive distributions associated with different tendon depths $a$ is quantified using the GP surrogates following Algorithm~\ref{alg:delta_overlap}. Two complementary statistics are employed. The first is a Min-$\Delta$ metric, which measures the smallest distinguishable difference in tendon depth $\Delta a$ such that the corresponding predictive distributions at a given sensor location remain separable under the propagated uncertainty. Smaller values of $\Delta a$ indicate higher sensitivity of the sensor to changes in $a$. The second is a PDF-overlap metric, which quantifies the degree of overlap between predictive probability density functions associated with different values of $a$; low overlap implies high discriminative power, while high overlap indicates limited ability to distinguish tendon breakage depths.

\begin{figure}[h!]
	\centering
	\includegraphics[width=\textwidth]{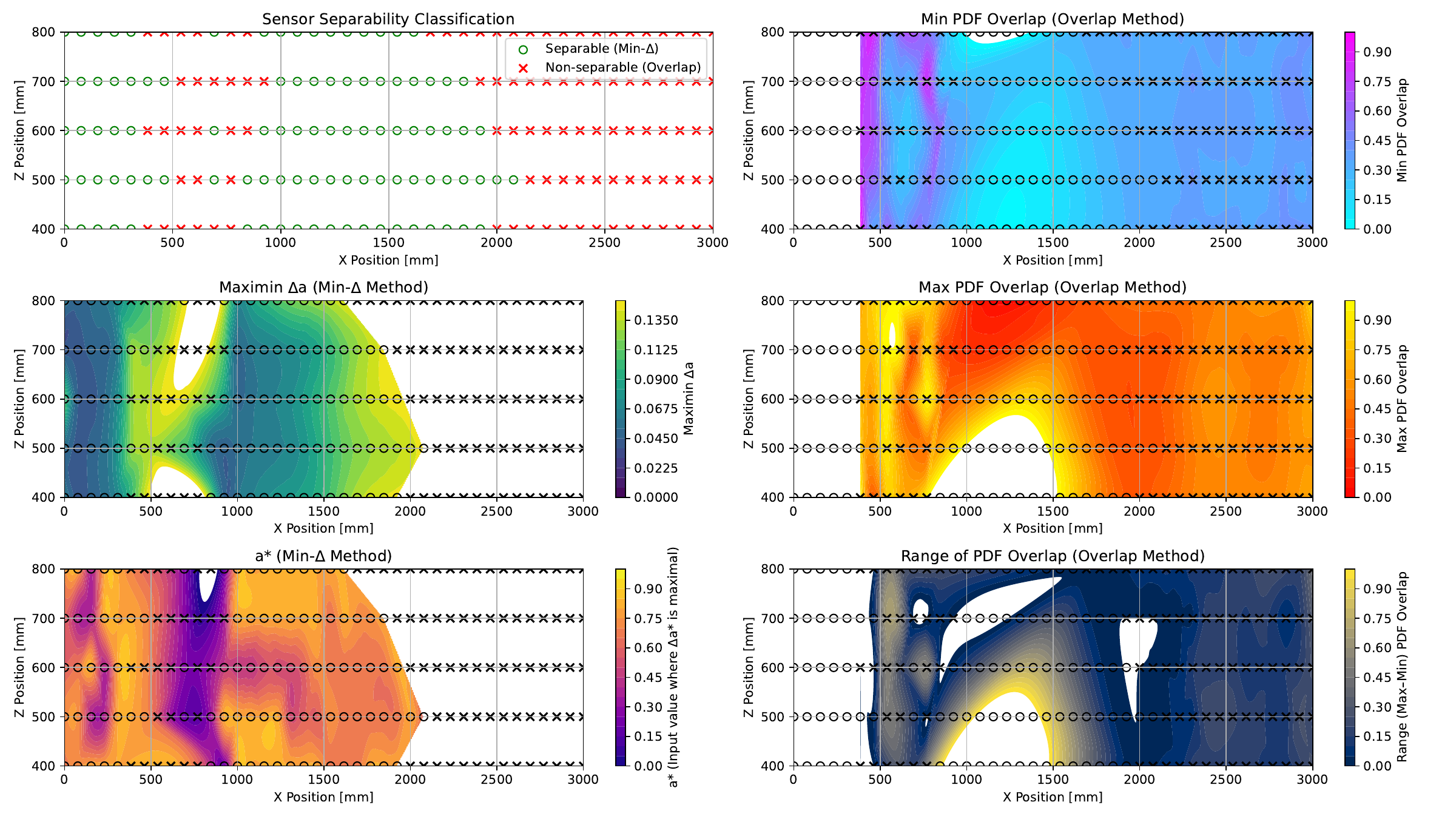}
	\caption{Comparison of sensor separability and overlap metrics for two classification approaches. Each subplot shows the spatial distribution of virtual sensors in the $x$--$z$ plane, with separable sensors ($\bm{\circ}$) identified by the Min-$\Delta$ method and non-separable sensors ($\bm{\times}$) identified by the PDF-overlap method. The position of the tendon breakage is at $x=0$ mm and $z=600$ mm. (Top left) Sensor separability classification map. (Middle left) Maximin $\Delta a$ field (Min-$\Delta$ method). (Bottom left) $a^*$ field indicating the tendon depth where $\Delta a^*$ is maximal. (Top right) Minimum PDF overlap field. (Middle right) Maximum PDF overlap field (Overlap method). (Bottom right) Range of PDF overlap summarizing overlap variability. Interpolation between applicable sensors is performed using cubic splines}
	\label{fig:pair_plot_mcmc}
\end{figure}

The optimization required to compute these statistics is performed using the Nelder–Mead method. The resulting spatial fields of separability and overlap are shown in Figure~\ref{fig:pair_plot_mcmc}. The top-left panel presents a classification of virtual sensor locations in the $x$-$z$ plane on the concrete surface ($y=t_w/2$) into separable ($\bm{\circ}$) and non-separable ($\bm{\times}$) regions based on the combined criteria of the two metrics. The middle-left panel shows the spatial distribution of the maximin $\Delta a$ values obtained with the Min-$\Delta$ method, highlighting regions where small changes in tendon depth can be reliably distinguished. The bottom-left panel reports the value $a^\ast$ at which $\Delta a$ is maximized, indicating which tendon depth is most influential at each sensor location. The top-right panel displays the minimum PDF overlap across all considered tendon depths, identifying sensor locations that provide the strongest discrimination. The middle-right panel shows the maximum overlap, revealing regions where predictive distributions remain highly overlapping regardless of $a$. The bottom-right panel summarizes the range of PDF overlap, providing a measure of variability in discriminative power across depth scenarios.

The results of the separability analysis reveal four distinct regions along the beam that correspond to different identifiability regimes. Close to the breakage point ($x<450$ mm), the predictive distributions for different tendon depths $a$ are well separated, indicating that the model response is highly sensitive to the breakage position and that the propagated material-form uncertainty does not obscure this sensitivity. In this region, the strain response is dominated by the local stress redistribution caused by the tendon break, which leads to measurable and distinguishable strain patterns across different tendon depths. 

Moving away from the breakage zone, a region of poor separability is observed (450 mm $<x<900$ mm). Here, the predictive distributions corresponding to different $a$ values exhibit significant overlap. This behaviour can be attributed to the interaction between the geometric eccentricity of the broken tendon, which induces a bending moment  around the vertical axis $z$ in the structure, and the uncertainty embedded in the stochastic Young’s modulus $\tilde{E}_{cm}$. The superposition of these two effects generates similar strain patterns for different tendon depths, thereby reducing the discriminatory power of the model in this intermediate region. 

Further along the beam (900 mm $<x<1900$ mm), the separability increases again, as the bending-induced strains stabilize and the relative effect of the tendon position on the overall strain field becomes more distinct. However, this trend does not persist indefinitely: far from the breakage point ($x>1900$ mm), the strain changes become too small to be distinguishable within the propagated uncertainty, and the predictive distributions once again overlap. This spatial pattern confirms that the propagation of the material-form uncertainty through the calibration framework provides valuable diagnostic information. It identifies regions where predictions are robust and distinguishable, and conversely, where model predictions should not be trusted for inverse inference. Consequently, this information can directly guide sensor placement strategies in practice, ensuring that measurements are taken in regions of maximal separability when the number of available sensors is limited. This assessment is specific to the calibrated model, a different set-up would generate other predictive distributions and affect the separability.

It should be noted that the present separability assessment is conducted on a single point-sensor basis, evaluating each spatial location independently. Considering multiple sensor locations simultaneously or a DFOS, and accounting for their spatial correlation, would enable a more comprehensive and tighter assessment of separability, potentially improving the identification of the breakage depth. This extension is left for future work. Additionally, the possible error introduced by transferring the calibrated parameters from the experimental setup to the upscaled model has not been explicitly quantified, as such transfer errors are inherently model-dependent and cannot be rigorously evaluated within the current framework.

\section{Conclusions}
\label{sec5}
This work presents a comprehensive, uncertainty-aware workflow for tendon breakage localization in pre-stressed concrete members, integrating laboratory-scale tendon breakage experiments and distributed fiber-optic sensor (DFOS) measurements, high-fidelity finite element modeling (FEM), surrogate modeling via Gaussian Processes (GPs), and Bayesian inference with embedded model-form uncertainty (MFU). The framework addresses the fundamental challenge that direct calibration on real structures is infeasible, as destructive tendon-breakage tests can hardly be performed in situ. By first calibrating models under controlled laboratory conditions and embedding uncertainties probabilistically, the workflow enables transferable, trustworthy predictions of tendon breakage locations and depths in realistic structural configurations such as bridges.

The starting point of the workflow was a detailed FEM representation of the tendon-breakage experiment, capable of reproducing the nonlinear structural response from re-anchoring and strain redistribution following a tendon breakage. To enable efficient probabilistic calibration, the computationally intensive FEM simulations were replaced by GP surrogate models that emulate the structural response across the parameter space while retaining accuracy in the regions most informed by the data. This surrogation step provided the foundation for the subsequent Bayesian calibration, in which the parameters governing material and bond behavior were inferred using the MCMC sampler. A key achievement of this study lies in the explicit quantification of model form uncertainty through parameter embedding. By interpreting the effective Young's modulus $E_{cm}$ as a stochastic variable with an inferred variance $\sigma_{E_{cm}}$, the calibration gained the ability to represent structural and modelling discrepancies probabilistically. This embedding not only improved the statistical consistency between model predictions and observations but also provided a physically interpretable mechanism for uncertainty propagation, thereby increasing the reliability and trustworthiness of the inferred parameters. Building upon the calibrated model, an influence analysis based on $\phi$-divergence (using Kullback-Leibler-divergence) was conducted to quantify the impact of specific data subsets on the posterior distributions. This diagnostic step allowed the identification of regions and parameters most responsible for model discrepancy, most notably, the sensitivity of the posterior to measurements near the tendon breakage. The insights gained from this analysis can directly guide the improvement of the model, targeting the area around the tendon break.

The final stage of the workflow demonstrated the transferability and predictive power of the calibrated and uncertainty-embedded model. By propagating the stochastic parameters to a realistic structural configuration, the study assessed the identifiability of an additional parameter, e.g. the depth of a broken tendon, through a separability analysis of the propagated predictive distributions. This analysis confirmed that the embedded MFU preserves sufficient separability near the breakage point to enable tendon-depth identification, while also revealing spatial regions where predictions are unreliable due to overlapping uncertainties. These findings provide a rational basis for optimizing sensor placement and for assessing confidence in inverse identification tasks. In summary, the developed framework achieves a seamless integration of experimental evidence, physics-based simulation, surrogate modeling, and probabilistic reasoning. The resulting methodology forms a comprehensive, data-informed approach for uncertainty-aware model calibration and validation in structural applications, and establishes a foundation for extending such techniques to large-scale engineering systems.

Although the proposed workflow demonstrates a reliable integration of experimental data, surrogate modelling, and Bayesian uncertainty embedding, several limitations must be acknowledged. The calibration strongly depends on the quality and representativeness of the experimental DFOS measurements, which are specific to the laboratory setup. Consequently, the inferred posterior distributions and embedded uncertainty reflect the conditions and boundary effects of that configuration. When transferring the parameters to different geometries or scales, these dependencies may not be fully preserved, introducing unquantifiable epistemic uncertainty. The use of Gaussian Process surrogates and the embedding of model form uncertainty through a single stochastic variable $\tilde{E}_{cm}$ constitute additional simplifications. While these approaches enable efficient calibration and propagation, they assume that the main discrepancies can be captured by the variability in the effective stiffness, neglecting spatial or multi-parameter sources of model error. Similarly, both the influence and separability analyses were performed under independence assumptions, i.e. considering each sensor or subset separately, thus not fully capturing correlations that could enhance discriminative power. These limitations highlight that, although the framework provides a systematic and trustworthy foundation for uncertainty-aware model calibration and interpretation, future developments should aim at multi-source uncertainty representation, adaptive surrogate refinement, and joint sensor analysis to further improve generalization and reliability.

Future developments of the proposed framework will focus on extending its capability to capture more complex and realistic sources of uncertainty and to enhance its diagnostic resolution. Incorporating correlations between sensor measurements and spatially distributed observations \citep{Mu2022} will enable multivariate separability and influence analyses, providing tighter and more reliable identification of parameter changes. Adaptive surrogate modeling strategies \citep{Semler2023}, such as active learning or multi-fidelity Gaussian Processes, could be employed to refine the representation of highly nonlinear responses while maintaining computational efficiency. Moreover, the extension of the embedding concept to multiple model form uncertainty parameters, potentially coupled with spatial variability fields, would allow a more physically consistent quantification of discrepancies \citep{Strong2015}. Finally, integrating experimental design optimization into the workflow could guide sensor placement and data acquisition strategies, maximizing the information gain for future structural monitoring or model updating campaigns \citep{Mello2024, PerezOrozco2025, Zhang2025, Bakeer2025}.

\paragraph*{Acknowledgments}
The authors made use of ChatGPT to assist with the drafting of this article for the improvement of text clarity and grammar. GPT-4 and GPT-5 were accessed from \url{https://chatgpt.com/} and used without modification between July 2025 and February 2026. All scientific content, ideas and interpretations originate from the authors. Claude Sonnet 4.5 was used through the Github Copilot interface for assisted programming in generation of the code required to obtain the results. All the content has been thoroughly reviewed by the authors.

\paragraph*{Funding Statement}
This research was supported by the German Research Foundation (DFG) - Project number 461030501 in the special focus program SPP 2388/1 ``Hundert plus – Verlängerung der Lebensdauer komplexer Baustrukturen durch intelligente Digitalisierung'' (Hundred Plus - Extending the service life of complex building structures through intelligent digitalisation) in the subprojects B04: Pattern detection of internal prestressing wire breaks on concrete surfaces (project number 501774158) and C07: Data driven model adaptation for identification of digital twins of bridges (project number 501811638).

\paragraph*{Competing Interests}
None

\paragraph*{Data Availability Statement}
Replication data and code can be found at Zenodo:\newline \verb+https://doi.org/10.5281/zenodo.18713387+.

\paragraph*{Ethical Standards}
The research meets all ethical guidelines, including adherence to the legal requirements of the study country.

\paragraph*{Author Contributions}
	Conceptualization: M.W.; D.S.; P.M.; J.F.U. Methodology: D.A.A.; A.P. Formal Analysis: D.A.A.; A.P. Funding acquisition: M.W.; D.S.; P.M.; J.F.U. Investigation: D.A.A.; A.P. Project Administration: M.W.; D.S.; P.M.; J.F.U. Resources: P.M.; J.F.U Software: D.A.A.; A.P. Data curation: D.A.A.; A.P. Data visualisation: D.A.A.; A.P. Writing original draft: D.A.A.; A.P. Supervision: M.W.; D.S.; P.M.; J.F.U. Review and Editing: M.W.; D.S.; P.M.; J.F.U. All authors approved the final submitted draft.

\paragraph*{Supplementary Material}
None

\appendix

\section{Clustering and Pruning of Walker Ensembles}
\label{ap:clustering-pruning}
After the burn-in phase of an ensemble MCMC sampler and at thee end of the full chain, we perform a postprocessing step that clusters walkers according to their average log-posterior values and prunes walkers belonging to low-probability clusters. The objective is to suppress walkers stuck in local minima in probability wells while preserving the largest coherent mode of the posterior. Our implementation is in line with \cite{Hou2012}, with some modifications to consider potential multiple low-probability clusters and to increase robustness against outliers.

Let the ensemble contain $N$ walkers and let $\{\ln \pi_{i,t}\}_{t=1}^{T}$ denote the log-posterior distribution of walker $i$ during the last $T = \lfloor \alpha\,T_{\mathrm{burn}}\rfloor$ steps of burn-in, where $\alpha\in(0,1)$ is a user-specified fraction, which in this paper will be 0.2. Define the mean terminal log-posterior
\begin{equation}
	\ell_i = \frac{1}{T}\sum_{t=1}^{T} \ln \pi_{i,t},
	\qquad i=1,\dots,N .
\end{equation}
Sort these values in ascending order $\ell_{(1)} \le \ell_{(2)} \le \dots \le \ell_{(N)}$, and denote the associated permutation of walker indices by $\sigma$. Define the first-order differences
\begin{equation}
	d_k = \ell_{(k+1)} - \ell_{(k)}, 
	\qquad k=1,\dots,N-1,
\end{equation}
and the median difference $\widetilde{d} \;=\; \mathrm{Median}\{d_k\}_{k=1}^{N-1}$. A jump is declared wherever
\begin{equation}
	d_k > \gamma\,\widetilde{d},
\end{equation}
where $\gamma>0$ is a user-specified jump factor that we will choose as 5.0. 

Let $J=\{k_1,k_2,\dots,k_M\}$ be the ordered list of detected jump locations. These induce $M+1$ disjoint clusters
\begin{equation}
	C_1 = \{1,\dots,k_1\},\ \
	C_2 = \{k_1+1,\dots,k_2\},\ \dots,\
	C_{M+1} = \{k_M+1,\dots,N\}.
\end{equation}
For each cluster $C_m$ with $m=1,...,M+1$, define its size	$S_m = |C_m|$. We select the \emph{largest} cluster index as
\begin{equation}
	m^\star = \arg\max_{m} S_m ,
\end{equation}
and define the acceptance threshold $\ell_{\mathrm{thr}} = \ell_{( \min C_{m^\star} )}$. All walkers satisfying $\ell_i \ge \ell_{\mathrm{thr}}$ are retained. Let $K$ be the set of retained indices and $P$ its complement.

For the next MCMC phase, let $\mathbf{x}_{i}^{\mathrm{last}}$ denote the last state of walker $i$ during burn-in. Construct an array $\mathbf{X}^{\mathrm{new}} \in \mathbb{R}^{N\times d}$, where $d$ is the parameter dimension, by
\begin{equation}
	\mathbf{X}^{\mathrm{new}}_{i} =
	\begin{cases}
		\mathbf{x}_{i}^{\mathrm{last}}, & i \in K, \\[4pt]
		\text{resampled point}, & i \in P.
	\end{cases}
\end{equation}
Resampling of pruned walkers is performed by convex mixing of two randomly drawn kept walkers
\begin{equation}
	\mathbf{X}^{\mathrm{new}}_{i}=w\, \mathbf{X}^{\mathrm{new}}_{a}+(1-w)\mathbf{X}^{\mathrm{new}}_{b},\qquad i\in P,
\end{equation}
where $a,b\in K$ are drawn without replacement and $w\sim \mathrm{Uniform}(0,1)$. This preserves the support of the dominant cluster while replenishing the ensemble with diversified points lying within the high-probability region.

At the end of the full chain, the same clustering formulation is reapplied to the final ensemble. In this second application, pruning is performed without resampling; we simply return the subset of walkers satisfying $\ell_i\ge\ell_{\mathrm{thr}}$ as the final reduced posterior ensemble.

\section{Stable estimation of $D_\phi(S)$ measures from posterior samples}
\subsection{Estimation of global influence}
\label{ap:influence_global}
Given a posterior sample $\left\lbrace\bm{\theta}^{(i)}\right\rbrace_{i=1}^{N} \sim \pi(\bm{\theta} \mid Y)$, we can estimate $D_\phi(S)$ from Equation \ref{eq:influence_global} empirically as
\begin{equation}
\widehat{D}_\phi(S)
= \log \left(\frac{1}{N}\sum_{i=1}^{N} \pi_S\left(\bm{\theta}^{(i)}\right)^{-1}\right)
+ \frac{1}{N}\sum_{i=1}^{N} \log \pi_S\left(\bm{\theta}^{(i)}\right).
\end{equation}
To prevent numerical underflow or overflow in evaluating likelihoods and ratios, all computations are performed in log-space. Let
\begin{equation}
\ell_i = \log \pi_S\left(\bm{\theta}^{(i)}\right) = \log \pi\left(Y \mid \bm{\theta}^{(i)}\right) - \log \pi\left(Y_{S^c} \mid \bm{\theta}^{(i)}\right).
\end{equation}
Define $M = \max\limits_i -\ell_i$ to stabilize exponentiation, and the scaled weights
\begin{equation}
\tilde{w}_i = \exp(-\ell_i - M) \quad \text{so that} \quad 0 < \tilde{w}_i \leq 1.
\end{equation}
Then, the log-space equivalent of the first term in $D_\phi(S)$ can be written using the log-sum-exp operation as
\begin{equation}
\log \mathbb{E}_{\bm{\theta} \mid Y}\left[\pi_S(\bm{\theta})^{-1}\right]
\approx -\log N + M + \log\left(\sum_{i=1}^{N} \tilde{w}_i\right)
= -\log N + \operatorname{logsumexp}(-\ell_i).
\end{equation}
Similarly, the second term is obtained as the sample mean of $\ell_i$ as
\begin{equation}
\mathbb{E}_{\bm{\theta} \mid Y}\left[\log \pi_S(\bm{\theta})\right]
\approx \frac{1}{N}\sum_{i=1}^{N} \ell_i.
\end{equation}
Finally, combining both terms gives the numerically stable estimator
\begin{equation}
\widehat{D}_\phi(S)
= \big[-\log N + \operatorname{logsumexp}(-\ell_i)\big]
+ \frac{1}{N}\sum_{i=1}^{N} \ell_i.
\end{equation}
\noindent This log-space formulation ensures stable and accurate computation of $D_\phi(S)$ even when the likelihood ratios $\pi_S(\bm{\theta})$ span several orders of magnitude, which is typical in high-dimensional or data-intensive Bayesian models.

\subsection{Estimation Marginal influence with KDE approach}
\label{ap:influence_kde}
Analogously, we approximate the two expectations of Equation \ref{eq:influence_kde} by empirical averages using posterior samples $\left\lbrace\theta_j^{(i)}\right\rbrace_{i=1}^N$ as
\begin{equation}
\widehat{D}_\phi^{(j)}(S)
= \log\left(\frac{1}{N}\sum_{k=1}^N \widehat{\pi}_S^{(j)}\left(\theta_j^{(k)}\right)^{-1}\right)
+ \frac{1}{N}\sum_{k=1}^N \log\left(\widehat{\pi}_S^{(j)}\left(\theta_j^{(k)}\right)\right).
\end{equation}
Because $w_i=\pi_S\left(\bm{\theta}^{(i)}\right)$ may be very small or large, we compute the kernel weights in log-space as well for numerical stability. Let $\ell_i=\log\pi_S\left(\bm{\theta}^{(i)}\right)$, $M=\max\limits_i -\ell_i$, and set scaled weights $\tilde w_i=\exp(-\ell_i-M)$. Then
\begin{equation}
\widehat{\pi}_S^{(j)}\left(\theta_j^{(k)}\right)
= \frac{\sum\limits_{i=1}^N \tilde w_i K_h\left(\theta_j^{(k)}-\theta_j^{(i)}\right)}
{\sum\limits_{i=1}^N K_h\left(\theta_j^{(k)}-\theta_j^{(i)}\right)}
\end{equation}
and the corresponding log-smoothed value is
\begin{equation}
\log\widehat{\pi}_S^{(j)}\left(\theta_j^{(k)}\right)= \log\left(\sum_{i=1}^N \tilde w_i K_h\left(\theta_j^{(k)}-\theta_j^{(i)}\right)\right)- \log\left(\sum_{i=1}^N K_h\left(\theta_j^{(k)}-\theta_j^{(i)}\right)\right) + M.
\end{equation}

\subsection{Estimation of marginal influence with Fixed Mean approach}
\label{ap:influence_fixed_mean}
The marginal influence of Equation \ref{eq:influence_fixed} is analogously approximated by samples ${\theta_j^{(i)}}$ as
\begin{equation}
\widehat{D}_\phi^{(j),(\text{fix})}(S)
= \log\left(\frac{1}{N}\sum_{i=1}^N \pi_S\left(\theta_j^{(i)},\bar{\theta}_{-j}\right)^{-1}\right)
+\frac{1}{N}\sum_{i=1}^N \log\left(\pi_S\left(\theta_j^{(i)},\bar{\theta}_{-j}\right)\right).
\end{equation}

\bibliographystyle{plainnat}
\bibliography{mybibliography}

@article{Abdelatif.2017,
 author = {Abdelatif, A. O. and Owen, J. S. and Hussein, M. F. M.},
 year = {2017},
 title = {Modeling and Parametric Study of the Reanchorage of Ruptured Tendons in Bonded Posttensioned Concrete},
 url = {https://ascelibrary.org/doi/full/10.1061/(asce)st.1943-541x.0001898},
 pages = {04017162},
 volume = {143},
 number = {12},
 issn = {0733-9445},
 journal = {Journal of Structural Engineering},
 doi = {10.1061/(ASCE)ST.1943-541X.0001898}
}

@Article{AndresArcones2024,
  author    = {Andrés Arcones, Daniel and Weiser, Martin and Koutsourelakis, Phaedon‐Stelios and Unger, Jörg F.},
  journal   = {Applied Stochastic Models in Business and Industry},
  title     = {Model Bias Identification for Bayesian Calibration of Stochastic Digital Twins of Bridges},
  year      = {2024},
  issn      = {1526-4025},
  month     = oct,
  copyright = {Creative Commons Attribution Non Commercial No Derivatives 4.0 International},
  doi       = {10.1002/asmb.2897},
  publisher = {Wiley},
  url       = {https://doi.org/10.1002/asmb.2897},
}

@Article{AndresArcones2024a,
  author        = {{Andrés Arcones}, Daniel and Weiser, Martin and Koutsourelakis, Phaedon-Stelios and Unger, Jörg F.},
  journal       = {Data-Centric Engineering},
  title         = {Embedded model form uncertainty quantification with measurement noise for Bayesian model calibration},
  year          = {2026},
  issn          = {2632-6736},
  month         = feb,
  volume        = {7},
  archiveprefix = {arXiv},
  doi           = {10.1017/dce.2025.10035},
  eprint        = {2410.12037},
  publisher     = {Cambridge University Press (CUP)},
  url           = {https://doi.org/10.1017/dce.2025.10035},
}

@article{Barrias.2016,
 author = {Barrias, Ant{\'o}nio and Casas, Joan R. and Villalba, Sergi},
 year = {2016},
 title = {A Review of Distributed Optical Fiber Sensors for Civil Engineering Applications},
 url = {https://www.mdpi.com/1424-8220/16/5/748},
 pages = {748},
 volume = {16},
 number = {5},
 issn = {1424-8220},
 journal = {Sensors},
 doi = {10.3390/s16050748}
}

@Article{Bayarri2007,
  author    = {Maria J Bayarri and James O Berger and Rui Paulo and Jerry Sacks and John A Cafeo and James Cavendish and Chin-Hsu Lin and Jian Tu},
  journal   = {Technometrics},
  title     = {A Framework for Validation of Computer Models},
  year      = {2007},
  month     = may,
  number    = {2},
  pages     = {138--154},
  volume    = {49},
  doi       = {10.1198/004017007000000092},
  publisher = {Informa {UK} Limited},
}

@article{Bergmeister.2015b,
 author = {Bergmeister, Konrad and Mark, Peter and {\"O}sterreicher, Michael and Sanio, David and Heek, Peter and Krawtschuk, Alexander and Strauss, Alfred and Ahrens, Mark Alexander},
 title = {Innovative Monitoringstrategien f{\"u}r Bestandsbauwerke},
 doi = {10.1002/9783433603406.ch7},
 journal = {BetonKalender},
 year = {2015}
}

@book{Birtel.2006,
 author = {Birtel, Veit and Mark, Peter},
 year = {2006},
 title = {Parameterised finite element modelling of RC beam shear failure}
}

@article{Briere.2013,
 author = {Briere, Vincent and Harries, Kent A. and Kasan, Jarret and Hager, Charles},
 year = {2013},
 title = {Dilation behavior of seven-wire prestressing strand -- The Hoyer effect},
 pages = {650--658},
 volume = {40},
 issn = {09500618},
 journal = {Construction and Building Materials},
 doi = {10.1016/j.conbuildmat.2012.11.064}
}

@article{Chapeleau.2021,
 author = {Chapeleau, Xavier and Bassil, Antoine},
 year = {2021},
 title = {A General Solution to Determine Strain Profile in the Core of Distributed Fiber Optic Sensors under Any Arbitrary Strain Fields},
 url = {https://www.mdpi.com/1424-8220/21/16/5423},
 pages = {5423},
 volume = {21},
 number = {16},
 issn = {1424-8220},
 journal = {Sensors},
 doi = {10.3390/s21165423}
}

@article{Clau.2021,
 author = {Clau{\ss}, Felix and Ahrens, Mark Alexander and Mark, Peter},
 year = {2021},
 title = {A comparative evaluation of strain measurement techniques in reinforced concrete structures},
 pages = {2992--3007},
 volume = {22},
 number = {5},
 issn = {1464-4177},
 journal = {Structural Concrete},
 doi = {10.1002/suco.202000706}
}

@misc{DassaultSystemesSimuliaCorporation.,
 author = {{Dassault Syst{\`e}mes Simulia Corporation}},
 title = {Abaqus Theory Guide}
}

@article{Farrar.2007,
 author = {Farrar, Charles R. and Worden, Keith},
 year = {2007},
 title = {An introduction to structural health monitoring},
 pages = {303--315},
 volume = {365},
 number = {1851},
 issn = {1364-503X},
 journal = {Philosophical Transactions of the Royal Society A: Mathematical, Physical and Engineering Sciences},
 doi = {10.1098/rsta.2006.1928}
}

@article{Hegger.2010,
 author = {Hegger, Josef and Bertram, Guido},
 year = {2010},
 title = {Verbundverhalten von vorgespannten Litzen in UHPC},
 pages = {379--389},
 volume = {105},
 number = {6},
 issn = {0005-9900},
 journal = {Beton- und Stahlbetonbau},
 doi = {10.1002/best.201000019}
}

@article{Herbers.2023,
 author = {Herbers, Max and Richter, Bertram and Gebauer, Daniel and Classen, Martin and Marx, Steffen},
 year = {2023},
 title = {Crack monitoring on concrete structures: Comparison of various distributed fiber optic sensors with digital image correlation method},
 pages = {6123--6140},
 volume = {24},
 number = {5},
 issn = {1464-4177},
 journal = {Structural Concrete},
 doi = {10.1002/suco.202300062}
}

@article{Hillerborg.1983,
 author = {Hillerborg, Arne},
 year = {1983},
 title = {Analysis of one single crack},
 url = {https://portal.research.lu.se/en/publications/analysis-of-one-single-crack},
 pages = {223--249},
 journal = {Fracture Mechanics of Concrete (Developments in civil engineering)}
}

@article{Janiak.2023,
 author = {Janiak, Till and Becks, Henrik and Camps, Benjamin and Classen, Martin and Hegger, Josef},
 year = {2023},
 title = {Evaluation of distributed fibre optic sensors in structural concrete},
 url = {https://link.springer.com/article/10.1617/s11527-023-02222-9},
 pages = {1--18},
 volume = {56},
 number = {9},
 issn = {1871-6873},
 journal = {Materials and Structures},
 doi = {10.1617/s11527-023-02222-9}
}

@article{Jankowiak.2005,
 author = {Jankowiak, Tomasz and Lodygowski, Tomasz},
 year = {2005},
 title = {Identification of parameters of concrete damage plasticity constitutive model},
 url = {https://yadda.icm.edu.pl/baztech/element/bwmeta1.element.baztech-article-bpp1-0059-0053},
 pages = {53--69},
 number = {6},
 journal = {Foundations of Civil and Environmental Engineering}
}

@Article{Kennedy2001,
  author     = {Marc C. Kennedy and Anthony O'Hagan},
  journal    = {Journal of the Royal Statistical Society: Series B (Statistical Methodology)},
  title      = {Bayesian calibration of computer models},
  year       = {2001},
  number     = {3},
  pages      = {425--464},
  volume     = {63},
  doi        = {10.1111/1467-9868.00294},
  publisher  = {Wiley},
}

@article{Konertz.2019,
 author = {Konertz, Dustin and L{\"o}schmann, Jens and Clau{\ss}, Felix and Mark, Peter},
 year = {2019},
 title = {Faseroptische Messung von Dehnungs- und Temperaturfeldern},
 pages = {292--300},
 volume = {94},
 number = {07-08},
 issn = {0005-6650},
 journal = {Bauingenieur},
 doi = {10.37544/0005-6650-2019-07-08-70}
}

@article{Kratzig.2004,
 author = {Kr{\"a}tzig, Wilfried B. and P{\"o}lling, Rainer},
 year = {2004},
 title = {An elasto-plastic damage model for reinforced concrete with minimum number of material parameters},
 pages = {1201--1215},
 volume = {82},
 number = {15-16},
 issn = {00457949},
 journal = {Computers {\&} Structures},
 doi = {10.1016/j.compstruc.2004.03.002}
}

@article{Lee.1998,
 author = {Lee, Jeeho and Fenves, Gregory L.},
 year = {1998},
 title = {Plastic-Damage Model for Cyclic Loading of Concrete Structures},
 pages = {892--900},
 volume = {124},
 number = {8},
 issn = {0733-9399},
 journal = {Journal of Engineering Mechanics},
 doi = {10.1061/(ASCE)0733-9399(1998)124:8(892)}
}

@article{Lubliner.1989,
 author = {Lubliner, Jacob and Oliver, Javier and Oller, Sergio. and O{\~n}ate, Eugenio},
 year = {1989},
 title = {A plastic-damage model for concrete},
 pages = {299--326},
 volume = {25},
 number = {3},
 issn = {00207683},
 journal = {International Journal of Solids and Structures},
 doi = {10.1016/0020-7683(89)90050-4}
}

@article{Paul2024,
	author    = {Paul, Aeneas and Sanio, David and Mark, Peter},
	journal   = {e-Journal of Nondestructive Testing},
	title     = {Monitoring tendon breaks in concrete structures at different depths using distributed fiber optical sensors},
	year      = {2024},
	issn      = {1435-4934},
	month     = jul,
	number    = {7},
	volume    = {29},
	doi       = {10.58286/29598},
	publisher = {NDT.net GmbH & Co. KG},
}

@article{Paul.2024b,
 author = {Paul, Aeneas and Sanio, David and Mark, Peter},
 year = {2024},
 title = {Detection of internal tendon breaks by fiber-optical measurements -- influence of the re-anchoring behavior},
 pages = {1287--1294},
 volume = {64},
 issn = {2452-3216},
 journal = {Procedia Structural Integrity},
 doi = {10.1016/j.prostr.2024.09.199}
}

@article{Paul.2025,
 author = {Paul, Aeneas and Sanio, David and Mark, Peter},
 year = {2025},
 title = {Detektion innerer Spannstahlbrüche durch faseroptische Messungen an Betonoberflächen},
 number = {120},
 issn = {1437-1006},
 journal = {Beton- und Stahlbetonbau},
 doi = {10.1002/best.202400100}
}

@article{Paul.2026,
 author = {Paul, Aeneas and Sanio, David and Mark, Peter},
 year = {2026},
 title = {Sensitivity Study on Accuracy and Relevant Parameters in Numerical Modeling of Internal Tendon Breaks},
 journal = {Structural Concrete (submitted)}
}

@article{Pirskawetz.2023,
 author = {Pirskawetz, Stephan Matthias and Schmidt, Sebastian},
 year = {2023},
 title = {Detection of wire breaks in prestressed concrete bridges by Acoustic Emission analysis},
 pages = {100151},
 volume = {14},
 issn = {26661659},
 journal = {Developments in the Built Environment},
 doi = {10.1016/j.dibe.2023.100151}
}

@Book{Rasmussen2006,
  author    = {Rasmussen, Carl Edward},
  publisher = {MIT Press},
  title     = {Gaussian processes for machine learning},
  year      = {2006},
  isbn      = {026218253X},
}

@article{Richter.2025,
 author = {Richter, Bertram and Will, Elias and Herbers, Max and Marx, Steffen},
 year = {2025},
 title = {Detection of prestressing wire breaks in post-tensioned concrete structures using distributed fiber optic strain sensing},
 issn = {2190-5452},
 journal = {Journal of Civil Structural Health Monitoring},
 doi = {10.1007/s13349-025-01018-5}
}

@Article{Sargsyan2019,
  author    = {Khachik Sargsyan and Xun Huan and Habib N. Najm},
  journal   = {International Journal for Uncertainty Quantification},
  title     = {Embedded model error representation for {Bayesian} model calibration},
  year      = {2019},
  number    = {4},
  pages     = {365--394},
  volume    = {9},
  comment   = {Overview on model bias methods in introduction!},
  doi       = {10.1615/int.j.uncertaintyquantification.2019027384},
  publisher = {Begell House},
}

@article{Samiec.2011,
 author = {Samiec, D.},
 year = {2011},
 title = {Verteilte faseroptische Temperatur- und Deh- nungsmessung mit sehr hoher Ortsaufl{\"o}sung},
 pages = {34--37},
 number = {6},
 journal = {Photonik}
}

@article{scikit-learn2011,
  title={Scikit-learn: Machine Learning in {P}ython},
  author={Pedregosa, F. and Varoquaux, G. and Gramfort, A. and Michel, V.
          and Thirion, B. and Grisel, O. and Blondel, M. and Prettenhofer, P.
          and Weiss, R. and Dubourg, V. and Vanderplas, J. and Passos, A. and
          Cournapeau, D. and Brucher, M. and Perrot, M. and Duchesnay, E.},
  journal={Journal of Machine Learning Research},
  volume={12},
  pages={2825--2830},
  year={2011}
}

@article{Sieradzki.1987,
 author = {Sieradzki, K. and Newman, R. C.},
 year = {1987},
 title = {Stress-corrosion cracking},
 pages = {1101--1113},
 volume = {48},
 number = {11},
 issn = {0022-3697},
 journal = {Journal of Physics and Chemistry of Solids},
 doi = {10.1016/0022-3697(87)90120-x}
}

@article{Speck.2019,
 author = {Speck, Kerstin and Vogdt, Fritz and Curbach, Manfred and Petryna, Yuri},
 year = {2019},
 title = {Faseroptische Sensoren zur kontinuierlichen Dehnungsmessung im Beton},
 pages = {160--167},
 volume = {114},
 number = {3},
 issn = {1437-1006},
 journal = {Beton- und Stahlbetonbau},
 doi = {10.1002/best.201800105}
}

@article{Strater.2024,
 author = {Str{\"a}ter, Noah and Clau{\ss}, Felix and Ahrens, Mark Alexander and Mark, Peter},
 year = {2024},
 title = {Detection of tendon breaks in prestressed concrete structures using coda wave interferometry},
 issn = {1464-4177},
 journal = {Structural Concrete},
 doi = {10.1002/suco.202400680}
}

@article{vanMeirvenne.2018,
 author = {{van Meirvenne}, Kizzy and de Corte, Wouter and Boel, Veerle and Taerwe, Luc},
 year = {2018},
 title = {Non-linear 3D finite element analysis of the anchorage zones of pretensioned concrete girders and experimental verification},
 pages = {764--779},
 volume = {172},
 issn = {0141-0296},
 journal = {Engineering Structures},
 doi = {10.1016/j.engstruct.2018.06.065}
}

@article{vanMier.1987,
 author = {{van Mier}, J. G.M. and Reinhard, H. W. and {van der Vlugt}, B. W.},
 year = {1987},
 title = {Ergebnisse dreiachsiger verformungsgesteuerter Belastungsversuche an Beton},
 pages = {353--361},
 number = {62},
 issn = {0005-6650},
 journal = {Bauingenieur}
}

@Article{Weiss1996a,
  author    = {Weiss, Robert},
  journal   = {Journal of the Royal Statistical Society Series B: Statistical Methodology},
  title     = {An Approach to Bayesian Sensitivity Analysis},
  year      = {1996},
  issn      = {1467-9868},
  month     = nov,
  number    = {4},
  pages     = {739--750},
  volume    = {58},
  doi       = {10.1111/j.2517-6161.1996.tb02112.x},
  publisher = {Oxford University Press (OUP)},
}

@Article{Zhu2012,
  author    = {Zhu, Hongtu and Ibrahim, Joseph G. and Cho, Hyunsoon and Tang, Niansheng},
  journal   = {Journal of Computational and Graphical Statistics},
  title     = {Bayesian Case Influence Measures for Statistical Models With Missing Data},
  year      = {2012},
  issn      = {1537-2715},
  month     = jan,
  number    = {1},
  pages     = {253--271},
  volume    = {21},
  doi       = {10.1198/jcgs.2011.10139},
  publisher = {Informa UK Limited},
}

@Book{Hastie2009,
  author    = {Hastie, Trevor and Tibshirani, Robert and Friedman, Jerome},
  publisher = {Springer New York},
  title     = {The Elements of Statistical Learning},
  year      = {2009},
  isbn      = {9780387848587},
  doi       = {10.1007/978-0-387-84858-7},
  issn      = {2197-568X},
  journal   = {Springer Series in Statistics},
}

@Book{Scott2008,
  author    = {Scott, David W.},
  publisher = {Wiley {\&} Sons, Incorporated, John},
  title     = {Multivariate Density Estimation Theory, Practice, and Visualization},
  year      = {2008},
  isbn      = {9780470316849},
  subtitle  = {Theory, Practice, and Visualization},
}

@Book{Forrester2008,
	author    = {Forrester, Alexander I. J. and Sóbester, András and Keane, Andy J.},
	publisher = {Wiley},
	title     = {Engineering Design via Surrogate Modelling: A Practical Guide},
	year      = {2008},
	isbn      = {9780470770801},
	month     = jul,
	doi       = {10.1002/9780470770801},
}

@Article{Xiu2002,
	author    = {Xiu, Dongbin and Karniadakis, George Em},
	journal   = {SIAM Journal on Scientific Computing},
	title     = {The Wiener--Askey Polynomial Chaos for Stochastic Differential Equations},
	year      = {2002},
	issn      = {1095-7197},
	month     = jan,
	number    = {2},
	pages     = {619--644},
	volume    = {24},
	doi       = {10.1137/s1064827501387826},
	publisher = {Society for Industrial & Applied Mathematics (SIAM)},
}

@Misc{Sudret2021,
	author    = {Bruno Sudret},
	title     = {Polynomial chaos expansions in 90 minutes},
	year      = {2021},
	copyright = {http://rightsstatements.org/page/InC-NC/1.0/},
	doi       = {10.3929/ETHZ-B-000508852},
	language  = {en},
	publisher = {ETH Zurich},
}

@InProceedings{Seiffert2019,
	author    = {Annemarie Seiffert and Andreas Jensen},
	booktitle = {5th International Conference on Smart Monitoring, Assessment and Rehabilitation of Civil Structures},
	title     = {A practical approach for modeling tendon and wire failures for model-based damage detection of prestressed concrete bridges},
	year      = {2019},
	address   = {Potsdam},
	month     = aug,
	publisher = {e-Journal of Non-destructive Testing},
	volume    = {25},
	url       = {https://www.ndt.net/?id=24910},
}

@Article{Ayoub2010,
	author    = {Ayoub, Ashraf and Filippou, Filip C.},
	journal   = {Journal of Structural Engineering},
	title     = {Finite-Element Model for Pretensioned Prestressed Concrete Girders},
	year      = {2010},
	issn      = {1943-541X},
	month     = apr,
	number    = {4},
	pages     = {401--409},
	volume    = {136},
	doi       = {10.1061/(asce)st.1943-541x.0000132},
	publisher = {American Society of Civil Engineers (ASCE)},
}

@Article{Sargsyan2015,
	author    = {Sargsyan, K. and Najm, H. N. and Ghanem, R.},
	journal   = {International Journal of Chemical Kinetics},
	title     = {On the Statistical Calibration of Physical Models},
	year      = {2015},
	issn      = {1097-4601},
	month     = feb,
	number    = {4},
	pages     = {246--276},
	volume    = {47},
	doi       = {10.1002/kin.20906},
	publisher = {Wiley},
}

@Article{Hou2012,
	author    = {Hou, Fengji and Goodman, Jonathan and Hogg, David W. and Weare, Jonathan and Schwab, Christian},
	journal   = {The Astrophysical Journal},
	title     = {An Affine-Invariant Sampler for Exoplanet Fitting and Discovery in Radial Velocity Data},
	year      = {2012},
	issn      = {1538-4357},
	month     = jan,
	number    = {2},
	pages     = {198},
	volume    = {745},
	doi       = {10.1088/0004-637x/745/2/198},
	publisher = {American Astronomical Society},
}

@Article{Watson1964,
  author  = {Geoffrey S. Watson},
  journal = {Sankhhyā: The Indian Journal of Statistics, Series A},
  title   = {Smooth Regression Analysis},
  year    = {1964},
  month   = dec,
  number  = {4},
  pages   = {359--372},
  volume  = {26},
}

@Article{Nadaraya1964,
  author    = {Nadaraya, E. A.},
  journal   = {Theory of Probability \& Its Applications},
  title     = {On Estimating Regression},
  year      = {1964},
  issn      = {1095-7219},
  month     = jan,
  number    = {1},
  pages     = {141--142},
  volume    = {9},
  doi       = {10.1137/1109020},
  publisher = {Society for Industrial & Applied Mathematics (SIAM)},
}

@Article{Perrin2025,
  author        = {Guillaume Perrin and Romain Jorge Do Marco and Christian Soize and Christine Funfschilling},
  title         = {Estimating Intractable Posterior Distributions through Gaussian Process regression and Metropolis-adjusted Langevin procedure},
  journal       = {arXiv},
  year          = {2025},
  month         = jun,
  archiveprefix = {arXiv},
  eprint        = {2506.13336},
  primaryclass  = {stat.ME},
}

@Article{Semler2023,
  author    = {Semler, Phillip and Weiser, Martin},
  journal   = {Inverse Problems},
  title     = {Adaptive Gaussian process regression for efficient building of surrogate models in inverse problems},
  year      = {2023},
  issn      = {1361-6420},
  month     = oct,
  number    = {12},
  pages     = {125003},
  volume    = {39},
  doi       = {10.1088/1361-6420/ad0028},
  publisher = {IOP Publishing},
}

@Article{Mu2022,
  author    = {Mu, He-Qing and Liang, Xin-Xiong and Shen, Ji-Hui and Zhang, Feng-Liang},
  journal   = {Sensors},
  title     = {Analysis of Structural Health Monitoring Data with Correlated Measurement Error by Bayesian System Identification: Theory and Application},
  year      = {2022},
  issn      = {1424-8220},
  month     = oct,
  number    = {20},
  pages     = {7981},
  volume    = {22},
  doi       = {10.3390/s22207981},
  publisher = {MDPI AG},
}

@Article{Strong2015,
  author    = {Mark Strong and Jeremy E. Oakley and Alan Brennan and Penny Breeze},
  journal   = {Medical Decision Making},
  title     = {Estimating the Expected Value of Sample Information Using the Probabilistic Sensitivity Analysis Sample},
  year      = {2015},
  month     = {mar},
  number    = {5},
  pages     = {570--583},
  volume    = {35},
  doi       = {10.1177/0272989x15575286},
  publisher = {{SAGE} Publications},
}

@Article{PerezOrozco2025,
  author    = {Pérez Orozco, José Andrés and Chamoin, Ludovic and Cortial, Julien and de Buhan, Maya and Soulier, Bruno},
  journal   = {Mechanical Systems and Signal Processing},
  title     = {Optimal placement of distributed optic fiber sensors with the modified Constitutive Relation Error},
  year      = {2025},
  issn      = {0888-3270},
  month     = sep,
  pages     = {113206},
  volume    = {238},
  doi       = {10.1016/j.ymssp.2025.113206},
  publisher = {Elsevier BV},
  ranking   = {rank5},
}

@Article{Zhang2025,
  author    = {Zhang, Chengwen and Chun, Qing and Lin, Yijie},
  journal   = {Measurement},
  title     = {Optimal sensor placement and structural health monitoring methods of ancient stone bridges based on an improved genetic algorithm: Taking Lugou Bridge as an example},
  year      = {2025},
  issn      = {0263-2241},
  month     = feb,
  pages     = {115680},
  volume    = {241},
  doi       = {10.1016/j.measurement.2024.115680},
  publisher = {Elsevier BV},
}

@Article{Becks2024,
  author    = {Becks, Henrik and Lippold, Lukas and Winkler, Paul and Moeller, Max and Rohrer, Maximilian and Leusmann, Thorsten and Anton, David and Sprenger, Bjarne and Kähler, Philipp and Rudenko, Iryna and Arcones, Daniel Andrés and Koutsourelakis, Phaedon-Stelios and Unger, Jörg F. and Weiser, Martin and Petryna, Yuri and Schnellenbach-Held, Martina and Lowke, Dirk and Wessels, Henning and Lenzen, Armin and Zabel, Volkmar and Könke, Carsten and Claßen, Martin and Hegger, Josef},
  journal   = {Bauingenieur},
  title     = {Neuartige Konzepte für die Zustandsüberwachung und -analyse von Brückenbauwerken – Einblicke in das Forschungsvorhaben SPP100+/Novel Concepts for the Condition Monitoring and Analysis of Bridge Structures – Insights into the SPP100+ Research Project},
  year      = {2024},
  issn      = {0005-6650},
  number    = {10},
  pages     = {327--338},
  volume    = {99},
  doi       = {10.37544/0005-6650-2024-10-63},
  publisher = {VDI Fachmedien GmbH and Co. KG},
}

@Article{Kang2025,
  author    = {Kang, Chongjie and Arcones, Daniel Andrés and Becks, Henrik and Beetz, Jakob and Blankenbach, Jörg and Claßen, Martin and Degener, Sebastian and Eisermann, Cedric and Göbels, Anne and Hegger, Josef and Hermann, Ralf and Kähler, Philipp and Peralta, Patricia and Petryna, Yuri and Schnellenbach‐Held, Martina and Schulz, Oliver and Smarsly, Kay and Fatih Sönmez, Mehmet and Sprenger, Bjarne and Unger, Jörg F. and Vassilev, Hristo and Weiser, Martin and Marx, Steffen},
  journal   = {Beton- und Stahlbetonbau},
  title     = {Intelligente digitale Methoden zur Verlängerung der Nutzungsdauer der Nibelungenbrücke},
  year      = {2025},
  issn      = {1437-1006},
  month     = dec,
  doi       = {10.1002/best.70070},
  publisher = {Wiley},
}

@Article{Herbers2024,
  author    = {Herbers, Max and Bartels, Jan‐Hauke and Richter, Bertram and Collin, Fabian and Ulbrich, Lisa and Al‐Zuriqat, Thamer and Chillón Geck, Carlos and Naraniecki, Hubert and Hahn, Oliver and Jesse, Frank and Smarsly, Kay and Marx, Steffen},
  journal   = {Beton- und Stahlbetonbau},
  title     = {openLAB – Eine Forschungsbrücke zur Entwicklung eines digitalen Brückenzwillings},
  year      = {2024},
  issn      = {1437-1006},
  month     = jan,
  doi       = {10.1002/best.202300094},
  publisher = {Wiley},
}

@Article{Mello2024,
  author    = {Mello, Felipe Martarella de Souza and Pereira, Joao Luiz Junho and Gomes, Guilherme Ferreira},
  journal   = {Journal of Sound and Vibration},
  title     = {Multi-objective sensor placement optimization in SHM systems with Kriging-based mode shape interpolation},
  year      = {2024},
  issn      = {0022-460X},
  month     = jan,
  pages     = {118050},
  volume    = {568},
  doi       = {10.1016/j.jsv.2023.118050},
  publisher = {Elsevier BV},
}

@Article{Bellman1970,
  author    = {Bellman, R. and Åström, K.J.},
  journal   = {Mathematical Biosciences},
  title     = {On structural identifiability},
  year      = {1970},
  issn      = {0025-5564},
  month     = apr,
  number    = {3–4},
  pages     = {329--339},
  volume    = {7},
  doi       = {10.1016/0025-5564(70)90132-x},
  publisher = {Elsevier BV},
}

@Article{Bakeer2025,
  author        = {Tammam Bakeer and Max Herbers and Steffen Marx},
  title         = {Sensor Informativeness, Identifiability, and Uncertainty in Bayesian Inverse Problems for Structural Health Monitoring},
  year          = {2025},
  month         = nov,
  journal       = {arXiv},
  archiveprefix = {arXiv},
  eprint        = {2511.16628},
  primaryclass  = {cs.CE},
}

@Article{Ostachowicz2019,
  author    = {Ostachowicz, Wieslaw and Soman, Rohan and Malinowski, Pawel},
  journal   = {Structural Health Monitoring},
  title     = {Optimization of sensor placement for structural health monitoring: a review},
  year      = {2019},
  issn      = {1741-3168},
  month     = jan,
  number    = {3},
  pages     = {963--988},
  volume    = {18},
  doi       = {10.1177/1475921719825601},
  publisher = {SAGE Publications},
}

\end{document}